\documentclass[12pt]{article}

\usepackage{indentfirst} 

\usepackage{amsfonts,amssymb,amsmath}
\usepackage{bm}          
\usepackage{cite}
\usepackage{url}
\usepackage{enumitem}
\usepackage{blkarray}
\usepackage[dvipsnames]{xcolor}
\usepackage{MnSymbol}
\usepackage{hyperref}
\usepackage{tensor}
\usepackage{arcs}
\usepackage{tikz-cd}
\usepackage{todonotes}
\usepackage[titletoc,title]{appendix}
\usepackage{todonotes}
\usepackage{tensor}
\usepackage{upgreek}
\usepackage{bbm}

\usepackage{etoolbox}

\gappto\normalsize{%
  \setlength{\abovedisplayskip}{5pt plus 2pt minus 3pt}%
  \setlength{\belowdisplayskip}{5pt plus 2pt minus 3pt}%
  \setlength{\abovedisplayshortskip}{0pt plus 2pt}%
  \setlength{\belowdisplayshortskip}{3pt plus 1pt minus 1pt}%
}

\usetikzlibrary{arrows}

\AtBeginDocument{}%

\newcommand{\corurl}{red}
\newcommand{\corcite}{ForestGreen}
\newcommand{\corlink}{blue}
\definecolor{dgreen}{cmyk}{1,0,1,0.25}
\newcommand{\dd}{\mathrm{d\!\raisebox{-0.05pt}{\scalebox{.8}[1.0]{I}}}}

\newcommand{\E}{\mathsf{E}}

\newcommand{\bd}{\mathrm{d}}
\newcommand{\iprod}{\raisebox{8pt}{\scalebox{1}[-2.5]{$\neg\,$}}}

\newcommand{\ee}{\mathsf{e}}

\newcommand{\ddd}{\mathrm{d}}
\newcommand{\FF}{\mathrm{F}}
\newcommand{\Aa}{\mathsf{A}}
\newcommand{\Ss}{\mathsf{S}}
\newcommand{\Gg}{\mathsf{g}}
\newcommand{\uu}{\mathsf{u}}

\newcommand{\g}{\gamma^{-1}}

\newcommand{\et}{{e_{\mathrm{t}}}}
\newcommand{\At}{{A_{\mathrm{t}}}}
\newcommand{\Ot}{{\Omega_{\mathrm{t}}}}
\newcommand{\taut}{{\tau_{\mathrm{t}}}}

\newcommand{\stackmark}[3]{%
  \mathbin{%
    \mathchoice%
      {\ooalign{$\displaystyle#1$\cr\hidewidth\raisebox{#3}{$\scriptstyle#2$}\hidewidth\cr}}%
      {\ooalign{$\textstyle#1$\cr\hidewidth\raisebox{#3}{$\scriptstyle#2$}\hidewidth\cr}}%
      {\ooalign{$\scriptstyle#1$\cr\hidewidth\raisebox{#3}{$\scriptscriptstyle#2$}\hidewidth\cr}}%
      {\ooalign{$\scriptscriptstyle#1$\cr\hidewidth\raisebox{#3}{$\scriptscriptstyle#2$}\hidewidth\cr}}%
  }%
}

\newcommand{\doublewedgebase}{%
  \raisebox{0ex}{\scalebox{0.75}{$\wedge$}}\mkern-10mu\wedge%
}

\newcommand{\wed}{\stackmark{\wedge}{\cdot}{-0.2ex}}

\newcommand{\wwed}{%
  \mathbin{%
    \mathchoice%
      {\ooalign{$\displaystyle\doublewedgebase$\cr\hidewidth\raisebox{-0.2ex}{\kern-0.1em$\scriptstyle\cdot$}\hidewidth\cr}}%
      {\ooalign{$\textstyle\doublewedgebase$\cr\hidewidth\raisebox{-0.2ex}{\kern-0.21em$\scriptstyle\cdot$}\hidewidth\cr}}%
      {\ooalign{$\scriptstyle\doublewedgebase$\cr\hidewidth\raisebox{-0.2ex}{\kern-0.05em$\scriptscriptstyle\cdot$}\hidewidth\cr}}%
      {\ooalign{$\scriptscriptstyle\doublewedgebase$\cr\hidewidth\raisebox{-0.2ex}{\kern-0.05em$\scriptscriptstyle\cdot$}\hidewidth\cr}}%
  }%
}

\newcommand{\iproddot}{%
  \mathbin{%
    \mathchoice%
      {\ooalign{$\displaystyle\iprod$\cr\hidewidth\raisebox{-0.8ex}{\kern-0.05em$\scriptstyle\cdot$}\hidewidth\cr}}%
      {\ooalign{$\textstyle\iprod$\cr\hidewidth\raisebox{-0.8ex}{\kern-0.05em$\scriptstyle\cdot$}\hidewidth\cr}}%
      {\ooalign{$\scriptstyle\iprod$\cr\hidewidth\raisebox{-0.6ex}{\kern-0.03em$\scriptscriptstyle\cdot$}\hidewidth\cr}}%
      {\ooalign{$\scriptscriptstyle\iprod$\cr\hidewidth\raisebox{-0.5ex}{\kern-0.03em$\scriptscriptstyle\cdot$}\hidewidth\cr}}%
  }%
}

\newcommand{\veee}{\stackmark{\vartriangle}{\cdot}{+0.2ex}}

\hypersetup{linktocpage,colorlinks,urlcolor=\corurl,citecolor=\corcite,linkcolor=\corlink,
pdftitle={Cartan geometry and the Hamiltonian analysis of gravitational actions in terms of forms},
pdfauthor={J.F. Barbero, J. Margalef-Bentabol, V. Varo, E.J.S. Villase\~nor}}

\numberwithin{equation}{section}  

\def\QED{{\boldmath$\rule{0.5em}{0.5em}$}}                                
\def\markatright#1{\leavevmode\unskip\nobreak\quad\hspace*{\fill}{#1}}    
\def\qed{\markatright{\QED}}                                              

\usepackage{titling}
\usepackage{authblk}
\title{Cartan geometry and the Hamiltonian analysis of gravitational actions in terms of forms}

\author[1]{J. Fernando Barbero G.}
\author[2]{Juan Margalef-Bentabol\,}
\author[3,4]{Valle Varo\,}
\author[3,4]{Eduardo J.S. Villase\~nor\,}
\affil[1]{\href{https://ror.org/05rtchs68}{Instituto de Estructura de la Materia}, IEM-CSIC, Serrano 123, 28006 Madrid, Spain}
\affil[2]{Department of Mathematics and Statistics \href{https://ror.org/0161xgx34}{Universit\'e de Montr\'eal}, Montr\'eal, Qu\'ebec, Canada}
\affil[3]{\href{https://ror.org/03ths8210}{Universidad Carlos III de Madrid}, Departamento de Matem\'aticas, Avenida de la Universidad, 30
 (edificio Sabatini), 28911, Legan\'es (Madrid), Espa\~na}
\affil[4]{Grupo de Teor\'{\i}as de Campos y F\'{\i}sica Estad\'{\i}stica. Instituto Gregorio Mill\'an (UC3M). Unidad Asociada al Instituto de Estructura de la Materia, CSIC}
\date{}                     
\begin{document}
	
\maketitle
\date{June 30, 2026}


\begin{abstract}
The purpose of the paper is to review the Hamiltonian formulation for several actions related to general relativity that can be written in terms of differential forms. As we will show the powerful ideas of Cartan geometry can be used to simplify the Hamiltonian analysis and arrive at a coherent and comprehensive geometrical understanding of it.
\end{abstract}

\tableofcontents


%
%
	
\section{Introduction}\label{sec_Introduction}

The Hamiltonian approach to the dynamics of gravitational theories has received a lot of attention since the initial work by Dirac \cite{Dirac1} and the subsequent development of the ADM formalism \cite{Arnowitt:1962,Arnowitt:1962hi}. One reason for this is that Hamiltonians can be used as a first step toward quantization. In fact, Loop Quantum Gravity (LQG) started as a novel Hamiltonian description of General Relativity (GR) formulated in terms of Yang-Mills variables instead of the traditional geometrodynamical ones \cite{Ashtekar:1986yd,Ashtekar:1987gu}.

All the known actions for GR are defined in terms of singular Lagrangians. The Hamiltonian analysis of this type of Lagrangians is not straightforward and requires some care. The first comprehensive approach to dealing with them was developed by Dirac \cite{Dirac2} who proposed a celebrated ``algorithm'' for this purpose. His approach relies on the identification of the so called \textit{primary constraints} and the introduction of a \textit{total Hamiltonian} $H_T$ involving arbitrary objects---multipliers in the usual parlance---that must be fixed by requiring that the dynamics given by $H_T$ be consistent. In the process of finding these multipliers by solving the relevant equations, extra  \textit{secondary constraints} may appear, in which case additional consistency checks are needed. Although the whole procedure may get quite involved, if followed to the letter, it provides a consistent way to arrive at the sought for Hamiltonian description. It is important to bear in mind that Dirac's method guarantees that this formulation describes \textit{precisely} the dynamics given by the Euler-Lagrange equations.

There are other possible ways to handle singular Lagrangian systems. An interesting one, formulated in a geometric language in stark contrast with that of Dirac (see, however, \cite{BarberoG:2019ajq}), was developed by Gotay, Nester and Hinds (GNH) \cite{Gotay1,Gotay2}. There the emphasis was on the geometrical meaning of the constraints and the consistency requirements, which were phrased as \textit{tangency conditions}. The most important difference between the Dirac and GNH points of view is that Dirac's method provides a description in the full phase space of the theory under consideration whereas the GNH approach gives a formulation on the primary constraint submanifold $\mathsf{M}_0$. A relevant consequence of working with these different descriptions is that, although we have the canonical symplectic form in phase space with a simple expression in terms of conjugate variables, on $\mathsf{M}_0$ we are usually forced to work with a complicated, presymplectic form. This notwithstanding, it is sometimes possible to find new variables in terms of which this presymplectic form is simple. In fact, the Ashtekar formulation for GR can by found in this way.

Despite the conceptual clarity that the GNH method provides, it must be said that it is almost as hard to implement as Dirac's algorithm. At this point, it is important to understand what the main sources of difficulties are in both approaches. In Dirac's method, the conditions that fix the value of the Dirac multipliers (collectively denoted as $u$) are linear, inhomogeneous equations of the form $A.u=B$ where the matrix $A$ generically depends on the canonical variables and, hence, has a rank that also depends on them. In practice, at least in the examples that we will consider here, the form of the solutions obtained by solving these equations directly is very complicated. This precludes an easy interpretation of their solutions, makes it difficult to simplify them by using the constraints and, as a consequence, to carry out some necessary consistency checks. Something similar happens if one uses the GNH method. In this case, the most difficult step is the resolution of the equations for the components of the Hamiltonian vector fields that, ultimately, contain the information on the dynamics of the system. Also, as in Dirac's case, if these solutions are not appropriately simplified by using the available constraints, checking the consistency requirements (i.e., the tangency of the Hamiltonian vector field to the full constraint submanifold) becomes a hard and unforgiving task.

In this paper we will consider a parametric family of actions for Euclidean and Lorentzian GR that includes the Cartan-Palatini and the Holst ones as well as their Carrollian ($c\to 0$) limit. These actions are written in terms of differential forms and lead to \textit{first order} field equations. These two facts suggest a different approach to find the corresponding Hamiltonian descriptions. On one hand, as many differential forms operators interact neatly with pullbacks, it is straightforward to rewrite the field equations in $3+1$ form once a suitable foliation of space time is introduced. On the other, the fact that the equations are first order naturally leads to ``projected equations'' in $3+1$ form, that must be the same as those obtained by the more complicated Dirac or GNH methods. For this reason, the main part of the paper will be a unified, concise, streamlined---but complete---derivation of the Hamiltonian description for all these models.

A very important tool for our analysis will be \textit{Cartan geometry} because it provides a conceptual framework that illuminates the meaning of the geometric objects in terms of which our actions are written (interesting discussions about the applications of Cartan geometry to the interpretation and understanding of gravity theories in three and four dimensions can be found in \cite{Wise:2009fu, Wise:2006sm}). It also tells us a big deal about the structure of the Hamiltonian vector fields and their interpretation, especially with regard to the symmetries of the dynamics. In fact, the insights gained by relying on Cartan geometry can also be used to simplify the computations if the Dirac or GNH methods are employed.

The paper is organized as follows. After this introduction, we review in Section \ref{sec_Cartan_geometry} the main geometric ideas that we use in the paper. Section \ref{sec_Actions} is devoted to discuss the family of actions that we study, their symmetries and the equations of motion in covariant (i.e. spacetime) form. The geometric ideas discussed in Section \ref{sec_Cartan_geometry} play a relevant role here. Section \ref{sec_Hamiltonians} is devoted to find the Hamiltonian formulation for these actions by introducing a spacetime foliation and an appropriate projection procedure. The implementation of the time gauge and the introduction of Ashtekar variables is the purpose of Section \ref{sec_time_gauge}. Section \ref{sec_DiracGNH} provides a discussion on how the ideas of Cartan geometry can be used to facilitate the computations if the Dirac or GNH approaches are followed. We will not give too many details but we will provide the crucial information to complete the Hamiltonian analysis in these frameworks. We end the paper with our conclusions and some comments in Section \ref{sec_Conclussions}. Auxiliary material has been compiled in the appendices, in particular, our notation and some useful mathematical results are explained in Appendix \ref{app_notation}. Some results on top-forms, frame dualities and the simplification of the field equations are given in Appendix \ref{app_math}.

%
%
\section{Cartan Geometry}{\label{sec_Cartan_geometry}}

\subsection{General properties of a Cartan geometry}

Let $G$ be a Lie group and $H \leq G$ a closed subgroup, with corresponding Lie algebras $\mathfrak{h} \leq \mathfrak{g}$. A Cartan geometry of type $(G,H)$ on a smooth manifold $M$ \cite{Wise:2006sm,sharpe2000differential}  is a principal $H$-bundle
\[
\pi_P : P \to M
\]
endowed with a $\mathfrak{g}$-valued one-form
\[
\omega \in \Omega^1(P,\mathfrak{g}),
\]
called the Cartan connection, satisfying the following axioms:
\begin{enumerate}
    \item [C1:] $H$-equivariance:
    \[
    (R_h)^*\omega = \operatorname{Ad}(h^{-1})\,\omega,
    \qquad \forall h\in H;
    \]

    \item[C2:] Reproduction of fundamental vector fields:
    \[
    \omega(X^*) = X,
    \qquad \forall X\in\mathfrak{h},
    \]
    where $X^*$ denotes the fundamental vector field on $P$ generated by $X$, namely
    \[
    X^*(p)=\left.\frac{\mathrm{d}}{\mathrm{d} t}\right|_{t=0} p\exp(tX);
    \]

    \item[C3:] Absolute parallelism: for every $p\in P$, the linear map
    \[
    \omega_p:T_pP\to\mathfrak{g}
    \]
    is an isomorphism with inverse $\omega^{-1}_p: \mathfrak{g}\to T_pP$. Notice that
     \[
    T_{ph} R_{h^{-1}}\circ \omega^{-1}_{ph} =   \omega^{-1}_p\circ \operatorname{Ad}(h)\,
    \qquad \forall h\in H\,,\quad \forall p \in P;
    \]
\end{enumerate}

The homogeneous model of a Cartan geometry of type $(G,H)$ is the canonical principal bundle
\[
H \rightarrow G \rightarrow G/H
\]
endowed with the left Maurer--Cartan form
\[
\omega_G\in\Omega^1(G,\mathfrak{g}).
\]
Thus, the homogeneous model of a Cartan geometry of type $(G,H)$ is precisely the corresponding Klein geometry \cite{sharpe2000differential}.

The Cartan connection $\omega$ of $(H \rightarrow P \rightarrow M,\omega)$ generalizes the Maurer--Cartan form on $G$ and, in particular, provides a natural parallelization of $P$
\[
TP \xrightarrow{\;\omega\;} P\times \mathfrak{g}.
\]
More explicitly,
\[
P\times \mathfrak{g}\ni (p,X)\mapsto \omega_p^{-1}(X)\in TP,
\qquad
TP\ni v\mapsto (\pi_P(v),\omega(v))\in P\times \mathfrak{g}.
\]
and then, the vertical subbundle $\text{Ver}_p P\subset TP $ is defined by
\[
\text{Ver}_p P = \omega_p^{-1}(\mathfrak{h})\,.
\]
Notice, however that, in general $\omega$ does not provide an $H$-invariant definition of $\text{Hor}_p P$ (this is only true for the reductive geometries discussed below).
Moreover, for each $X\in\mathfrak{g}$, one defines the associated constant vector field
\[
\omega^{-1}(X)\in \mathfrak{X}(P)
\]
by
\[
p\mapsto \omega_p^{-1}(X).
\]
This vector field satisfies
\[
\omega\bigl(\omega^{-1}(X)\bigr)=X,
\]
where $X$ is regarded as a constant $\mathfrak{g}$-valued function on $P$,  and
\[
(R_h)_*\omega^{-1}(X)=\omega^{-1}(\operatorname{Ad}(h^{-1}
)X)\,.
\] Notice that, when $X\in\mathfrak{h}$, $\omega^{-1}(X)=X^*$ are just the fundamental (vertical) vector fields on $P$.

In the homogeneous model
\[
(H\rightarrow G\rightarrow G/H,\omega_G),
\]
one recovers
\[
\omega_G^{-1}(X)=L_X,
\]
where $L_X$ denotes the left-invariant vector field on $G$ generated by $X$.

Two $(G,H)$-Cartan geometries $(H\rightarrow P_1\rightarrow M_1,\omega_1)$ and  $(H\rightarrow P_2\rightarrow M_2,\omega_2)$  are isomorphic if and only if there exists a diffeomorphism $f:M_1\rightarrow M_2$ covered by a $H$-bundle isomorphism $\tilde{f}:P_1 \rightarrow P_2$ such that $\tilde{f}^*\omega_2=\omega_1$.

A $(G,H)$-Cartan geometry is called reductive if there exists a $\mathrm{Ad}$ $H$-module decomposition
\[
\mathfrak{g}=\mathfrak{h}\oplus\mathfrak{m}
\]
i.e., $\operatorname{Ad}(h)(\mathfrak{m}) \subseteq \mathfrak{m}$ for all $h \in H$ \cite{eschenburg1997lecture}. For reductive geometries, due to the Ad(H)-invariance,
\[
\mathrm{Ho}r_p(P)=\omega_p^{-1}(\mathfrak{m})\,.
\]
At the infinitesimal level, this implies the bracket relations $[\mathfrak{h}, \mathfrak{h}] \subseteq \mathfrak{h}$ and $[\mathfrak{h}, \mathfrak{m}] \subseteq \mathfrak{m}$. If, in addition, the reductive geometry satisfies
\[
[\mathfrak{h},\mathfrak{h}]\subset \mathfrak{h}\,,\quad
[\mathfrak{h},\mathfrak{m}]\subset \mathfrak{m}\,,\quad [\mathfrak{m},\mathfrak{m}]\subset \mathfrak{h}
\]
then it is called \textit{symmetric}. Many interesting examples in relativity---though not all---are symmetric. In the symmetric case, the linear map $\theta: \mathfrak{g} \to \mathfrak{g}$ defined by
\[
\theta(X_{\mathfrak{h}} + X_{\mathfrak{m}}) = X_{\mathfrak{h}} - X_{\mathfrak{m}},
\]
for $X_{\mathfrak{h}} \in \mathfrak{h}$ and $X_{\mathfrak{m}} \in \mathfrak{m}$, is a Lie algebra involution of $\mathfrak{g}$. One has $\mathfrak{h} = \{X \in \mathfrak{g} : \theta(X) = X\}$ and $\mathfrak{m} = \{X \in \mathfrak{g} : \theta(X) = -X\}$. Under mild assumptions on $(G,H)$, it can be shown that $\theta = T_e\sigma$, where $\sigma$ is an involutive automorphism of order two of $G$, although typically not an inner automorphism, as illustrated in the examples given below. In any case, the infinitesimal definition in the Lie algebra recovers the standard geometric definition.

\subsection{Cartan geometry in GR}

We discuss now the Cartan geometries  most commonly used in GR. Let $1+n$ be the spacetime dimension and $\Lambda$ the value of the cosmological constant. For each \(\kappa=\frac{2\Lambda }{(n-1)(n-2)}\in\mathbb{R}\) let us define the Lie group $G_\kappa$ by
\begin{equation*}
G_\kappa = \begin{cases}
O^{++}(1,n+1) & \text{if } \kappa > 0, \\
O^{++}(1,n) \ltimes\mathbb{R}^{1,n} & \text{if } \kappa = 0, \\
O^{++}(2,n) & \text{if } \kappa < 0,
\end{cases}
\end{equation*}
and $H_\kappa\leq G_\kappa$ with $H_\kappa\cong O^{++}(1,n)$, embedded in the natural way dictated by the algebra described below. Its Lie algebra $\mathfrak{g}_{\kappa}=\mathfrak{h}\oplus \mathfrak{m}_\kappa$, where $\mathfrak{h}=\mathfrak{o}(1,n)$, can be described in terms of matrices of the form
\begin{align*}
\mathfrak{g}_\kappa =\mathfrak{h}\oplus \mathfrak{m}_\kappa \ni \begin{pmatrix}
0 & 0 \\
0 & X
\end{pmatrix} + \begin{pmatrix}
0 & -\kappa x^t\eta \\
x & 0
\end{pmatrix},\quad X\in \mathfrak{o}(1,n),\quad x\in \mathbb{R}^{1,n},
\end{align*}
with the Lie bracket
\begin{align*}
\left[\begin{pmatrix}
0 & -\kappa x_1^t\eta \\
x_1 & X_1
\end{pmatrix}, \begin{pmatrix}
0 & -\kappa x_2^t\eta \\
x_2 & X_2
\end{pmatrix}\right] = \begin{pmatrix}
0 & -\kappa (X_1 x_2- X_2 x_1)^t\eta \\
X_1x_2-X_2x_1 & [X_1,X_2]-\kappa(x_1x_2^t-x_2x_1^t)\eta
\end{pmatrix}.
\end{align*}
As $\operatorname{Ad}(H)$-modules, all $\mathfrak{g}_\kappa$ are isomorphic (they are mutant Cartan geometries). As Lie algebras, there are three isomorphism classes corresponding to $\kappa<0$, $\kappa=0$, and $\kappa>0$. Meanwhile, as Cartan geometries, it is easy to show that the $(G_\kappa,H)$ for different values of $\kappa$ are inequivalent.

A representation for the Cartan connection (as an element of $\mathfrak{g}_{\kappa}$) is given by
\begin{align*}
\omega_{\kappa}= \begin{pmatrix}
0 & -\kappa e^t \eta \\
e & A
\end{pmatrix},
\end{align*}
where
\begin{align*}
(\omega_\kappa)_{\mathfrak{h}}:= \begin{pmatrix}
0 & 0 \\
0 & A
\end{pmatrix} \in \Omega^1(P,\mathfrak{h}),\quad
(\omega_\kappa)_{\mathfrak{m}_\kappa} := \begin{pmatrix}
0 & -\kappa e^t \eta \\
e  & 0
\end{pmatrix} \in \Omega^1(P,\mathfrak{m}_\kappa).
\end{align*}
The curvature of the Cartan connection is
\begin{align*}
F_{\omega_\kappa}
& = \mathrm{d}\omega_{\kappa} + \frac{1}{2} [\omega_{\kappa}, \omega_{\kappa}]
= \mathrm{d}\omega_{\kappa} + \omega_{\kappa} \wedge \omega_{\kappa}
= \begin{pmatrix}
0 & -\kappa (\mathrm{d}_A e)^t \eta \\
\mathrm{d}_A e & F_A - \kappa (e \wedge e^t)\eta
\end{pmatrix}.
\end{align*}
The projections $(F_{\omega_{\kappa}})_{\mathfrak{m}_\kappa}$ and $(F_{\omega_{\kappa}})_{\mathfrak{h}}$ correspond, respectively, to the torsion and the usual curvature of the geometry. The structural properties of the Maurer-Cartan form imply that the homogeneous model for a Cartan geometry is Cartan-flat, i.e., $F_{\omega_{\kappa}} = 0$. Hence,
$$\mathrm{d}_A e = 0,\qquad F_A = \kappa (e \wedge e^t)\eta.$$
Cartan-flatness thus forces $A$ to be the principal Levi-Civita connection associated with the coframe $e$, and the Riemann tensor $F_A$ to take the form $\kappa (e \wedge e^t)\eta$. Consequently, it describes a spacetime of constant sectional curvature $\kappa$, i.e.,
$$\langle F_A(u,v)u,v\rangle = \kappa \big(\langle u,u \rangle\langle v,v \rangle- \langle u,v \rangle^2\big).$$
When $\kappa = 0$, the homogeneous space $G_0/H$ is simply the Minkowski spacetime $\mathbb{R}^{1,n}$, whereas for $\kappa \neq 0$ the geometry is not Riemann-flat. Specifically, $\kappa > 0$ corresponds to de Sitter spacetime $G_\kappa/H$, with topology $\mathbb{R}\times \mathbb{S}^n$, and $\kappa < 0$ to anti-de Sitter spacetime with topology $\mathbb{S}^1\times \mathbb{R}^n$; that is
\begin{align*}
(H\rightarrow G_\kappa\rightarrow G_\kappa/H,\quad w_{G_\kappa})
\end{align*}
and
\begin{align*}
 (H\rightarrow FO^{++}(M_\kappa)\rightarrow  M_\kappa  \,,\quad w_{\kappa}^{\textrm{PLC}})\,.
\end{align*}
Notice that if one requires the homogeneous model for $\kappa<0$ to be simply connected and to avoid closed causal curves, $G_\kappa$ must be replaced by $\widetilde{O^{++}}(2,n)$ (a non-matrix Lie group).  Notice also that, in practice, these equations are not imposed by hand but arise naturally from a variational principle.

\subsection{Infinitesimal version on $\mathrm{Aut}(P)$ and $\mathrm{Gauge}(P)$}

Given a principal $H$-bundle
$$
H\to P\stackrel{\pi}{\to} M,
$$
its automorphism group $\mathrm{Aut}(P)$ consists of all $H$-equivariant diffeomorphisms of $P$. In other words, the diffeomorphisms that move $H$-orbits to $H$-orbits. Since $M$ can be understood as the space of orbits, by taking a representative of each orbit, an automorphism $\Phi$ induces  a diffeomorphism $f_\Phi$ on the base $M$. It is given by
$$
\pi\circ\Phi=f_\Phi\circ\pi.
$$
We have then a homomorphism $\pi_*:\mathrm{Aut}(P)\rightarrow\mathrm{Diff}(M)$, with $\pi_*(\Phi):=f_\Phi$, whose image is the subgroup $\mathrm{Diff}_P(M)\subseteq\mathrm{Diff}(M)$ consisting of those diffeomorphisms of the base which admit an $H$-equivariant lift to an automorphism of $P$. Equivalently,
$$
\mathrm{Diff}_P(M)
=
\{f\in\mathrm{Diff}(M)\mid f^*P\cong P\},
$$
where the isomorphism is understood as an isomorphism of principal $H$-bundles. Thus, the condition $f^*P\cong P$ is precisely the condition for the existence of a lift of $f$ to an automorphism of $P$, i.e., $f=f_\Phi$ for some $\Phi\in\mathrm{Aut}(P)$. Clearly, such a lift is not unique: any two lifts differ by a vertical automorphism i.e. an automorphism $\Psi$ that ``shuffles'' the points within each orbit, in other words, such that $f_\Psi=\mathrm{Id}$. The group of vertical automorphism is also known as the gauge group
$$
\mathrm{Gauge}(P)
=
\ker\bigl(\pi_*:\mathrm{Aut}(P)\to\mathrm{Diff}(M)\bigr)
$$
which is the normal subgroup of $\mathrm{Aut}(P)$ consisting of those automorphisms covering the identity on $M$. We thus obtain the short exact sequence
$$
1\longrightarrow\mathrm{Gauge}(P)
\longrightarrow\mathrm{Aut}(P)
\stackrel{\pi_*}{\longrightarrow}
\mathrm{Diff}_P(M)
\longrightarrow1.
$$
Let $(H\to P\stackrel{\pi}{\to}M,\omega)$ be a reductive Cartan geometry of type $(G,H)$. Let $X\in\mathfrak X(P)$ be an $H$-invariant vector field, i.e.
$$
(R_{h})_*X=X\qquad \forall h\in H.
$$
Such a vector field projects to a vector field $\bar X\in\mathfrak X(M)$ satisfying
$$
\pi_*X=\bar X\circ\pi.
$$
At the level of the underlying principal bundle, $X$ is the infinitesimal generator of a one-parameter family of principal bundle automorphisms. It is an infinitesimal automorphism of the Cartan geometry itself if, in addition, it preserves the Cartan connection, namely if
$$
\pounds_X\omega=0.
$$

The Cartan connection $\omega\in\Omega^1(P,\mathfrak g)$ identifies $T_pP\cong\mathfrak g$ at every $p\in P$ and therefore determines the $\mathfrak g$-valued function
$$
\lambda:=X\iprod\omega=\omega(X).
$$
Since $X$ is $H$-invariant and $\omega$ is $H$-equivariant, $\lambda$ is an $H$-equivariant function from $P$ to $\mathfrak g$. Thus, $\lambda$ provides the representation of the vector field $X$ in the Cartan-geometric model.

Using Cartan's formula,
$$
\pounds_X\omega=X\iprod \mathrm{d}\omega+\mathrm{d}(X\iprod\omega),
$$
together with the definition of the Cartan curvature
$$
F_\omega=\mathrm{d}\omega+\frac{1}{2}[\omega,\omega],
$$
the infinitesimal variation of the Cartan connection induced by $X$ can be written as
$$
\pounds_X\omega=\mathrm{d}_\omega\lambda+X\iprod F_\omega,
$$
where $\mathrm{d}_\omega\lambda=\mathrm{d}\lambda+[\omega,\lambda]$ denotes the covariant exterior derivative of the $\mathfrak g$-valued function $\lambda$ with respect to the Cartan connection $\omega$.

Consequently, the condition for $X$ to be an infinitesimal automorphism of the Cartan geometry is
$$
\mathrm{d}_\omega(X\iprod \omega)+X\iprod F_\omega=0.
$$

For a reductive Cartan geometry, the Lie algebra of $G$ admits a vector-space decomposition
$$
\mathfrak g=\mathfrak h\oplus\mathfrak m,
\qquad
[\mathfrak h,\mathfrak m]\subseteq\mathfrak m.
$$
This decomposition allows us to split both the Cartan connection and its curvature into their $\mathfrak h$- and $\mathfrak m$-components. In particular,
$$
\omega=\omega_\mathfrak h+\omega_\mathfrak m,
$$
where $\omega_\mathfrak h\in\Omega^1(P,\mathfrak h)$ is a principal $H$-connection on $P$, while
$\omega_\mathfrak m\in\Omega^1(P,\mathfrak m)$ is the solder form (or coframe).

The Cartan curvature decomposes accordingly as
$$
F_\omega=(F_\omega)_\mathfrak h+(F_\omega)_\mathfrak m,
$$
with
$$
(F_\omega)_\mathfrak h
=
F_{\omega_\mathfrak h}
+\frac{1}{2}[\omega_\mathfrak m,\omega_\mathfrak m]_\mathfrak h
\in\Omega^2(P,\mathfrak h).
$$
We refer to $(F_\omega)_\mathfrak h$ as the $\mathfrak h$-component of the Cartan curvature, or simply as the $H$-curvature. Its geometric meaning becomes particularly transparent when compared with the homogeneous model $G/H$. Indeed, for the canonical Cartan geometry on $G/H$, the Cartan connection is induced by the Maurer--Cartan form on $G$, and its full Cartan curvature vanishes. Consequently, its $\mathfrak h$-component satisfies
$$
F_{\omega_\mathfrak h}
+
\frac{1}{2}[\omega_\mathfrak m,\omega_\mathfrak m]_\mathfrak h
=0.
$$
Thus, $(F_\omega)_\mathfrak h$ measures the deviation of the $\mathfrak h$-curvature of the given Cartan geometry from the curvature determined by the homogeneous model $G/H$.

The $\mathfrak m$-component is
$$
(F_\omega)_\mathfrak m
=
\mathrm{d}_{\omega_\mathfrak h}\omega_\mathfrak m
+\frac{1}{2}[\omega_\mathfrak m,\omega_\mathfrak m]_\mathfrak m
\in\Omega^2(P,\mathfrak m),
$$
and is the torsion of the Cartan geometry. For the homogeneous model, the $\mathfrak m$-component of the Maurer--Cartan equation gives
$$
\mathrm{d}_{\omega_\mathfrak h}\omega_\mathfrak m
=
-\frac{1}{2}[\omega_\mathfrak m,\omega_\mathfrak m]_\mathfrak m.
$$
This motivates interpreting
$$
-\frac{1}{2}[\omega_\mathfrak m,\omega_\mathfrak m]_\mathfrak m
$$
as the intrinsic torsion of the reductive model $G/H$. Accordingly, the torsion $(F_\omega)_\mathfrak m$ measures the deviation of the covariant exterior derivative $\mathrm{d}_{\omega_\mathfrak h}\omega_\mathfrak m$ from its value in the homogeneous model. In particular, for a symmetric model,
$$
[\mathfrak m,\mathfrak m]\subseteq\mathfrak h,
$$
and hence the intrinsic torsion vanishes.

We may similarly decompose the function $\lambda\in C^\infty(P,\mathfrak g)$ as
$$
\lambda=\lambda_\mathfrak h+\lambda_\mathfrak m.
$$
Its two components are given by
$$
\lambda_\mathfrak h
=
X\iprod\omega_\mathfrak h
=
\omega_\mathfrak h(X)
\in C^\infty(P,\mathfrak h)
$$
and
$$
\lambda_\mathfrak m
=
X\iprod\omega_\mathfrak m
=
\omega_\mathfrak m(X)
\in C^\infty(P,\mathfrak m).
$$
The $\mathfrak h$-component $\lambda_\mathfrak h$ encodes the vertical part of $X$. The geometric meaning of $\lambda_\mathfrak m$ is slightly different. Since $\omega_\mathfrak m$ is the solder form, it induces the canonical identification
$$
TM\cong P\times_H\mathfrak m.
$$
Under this identification, the $H$-equivariant function $\lambda_\mathfrak m$ corresponds precisely to the vector field
$$
\bar X=\pi_*X
$$
on the base. Thus, $\lambda_\mathfrak m$ represents the infinitesimal displacement induced by $X$ on $M$. It should not, however, be interpreted as a horizontal component of $X$ in the usual connection-theoretic sense: the decomposition
$$
\lambda=\lambda_\mathfrak h+\lambda_\mathfrak m
$$
is induced by the reductive decomposition of $\mathfrak g$, rather than by a choice of horizontal distribution on $P$.

Projecting the general identity
$$
\pounds_X\omega=X\iprod F_\omega+\mathrm{d}_\omega\lambda
$$
onto $\mathfrak m$ and $\mathfrak h$ gives the corresponding variations of the two components of the Cartan connection. For the solder form, one obtains
$$
\pounds_X\omega_\mathfrak m
=
\mathrm{d}_{\omega_\mathfrak h}\lambda_\mathfrak m
+
[\omega_\mathfrak m,\lambda_\mathfrak h]
+
[\lambda_\mathfrak m,\omega_\mathfrak m]_\mathfrak m
+
X\iprod(F_\omega)_\mathfrak m,
$$
where
$$
\mathrm{d}_{\omega_\mathfrak h}\lambda_\mathfrak m
:=
\mathrm{d}\lambda_\mathfrak m
+
[\omega_\mathfrak h,\lambda_\mathfrak m]
$$
is the covariant exterior derivative associated with the principal connection $\omega_\mathfrak h$. The term
$$
[\lambda_\mathfrak m,\omega_\mathfrak m]_\mathfrak m
$$
denotes the $\mathfrak m$-projection of the Lie bracket in $\mathfrak g$. If the geometry is modelled on a symmetric space, so that
$$
[\mathfrak m,\mathfrak m]\subseteq\mathfrak h,
$$
this term vanishes.

Similarly, the variation of the $H$-connection is
$$
\pounds_X\omega_\mathfrak h
=
\mathrm{d}_{\omega_\mathfrak h}\lambda_\mathfrak h
+
[\lambda_\mathfrak m,\omega_\mathfrak m]_\mathfrak h
+
X\iprod(F_\omega)_\mathfrak h.
$$
It is useful to examine the geometric interpretation of these two identities in several particularly relevant special cases.
First, $X$ is called a Killing--Cartan vector field if the corresponding infinitesimal automorphism preserves the Cartan connection, i.e.
$$
\pounds_X\omega=0\quad \Leftrightarrow \quad\begin{array}{l}\pounds_X\omega_\mathfrak m=0,\\\pounds_X\omega_\mathfrak h=0.\end{array}
$$
In terms of the function $\lambda$, the Killing--Cartan condition is therefore
$$
\mathrm{d}_\omega\lambda+X\iprod F_\omega=0.
$$

Pure gauge transformations are recovered at the infinitesimal level by setting
$$
\lambda_\mathfrak m=0.
$$
Since $\omega$ is a pointwise isomorphism, this condition is equivalent to $X$ being vertical. The above identities then reduce to
$$
\pounds_X\omega_\mathfrak m
=
[\omega_\mathfrak m,\lambda_\mathfrak h],
\qquad
\pounds_X\omega_\mathfrak h
=
\mathrm{d}_{\omega_\mathfrak h}\lambda_\mathfrak h.
$$
Thus, under a vertical infinitesimal automorphism, the solder form $\omega_\mathfrak m$ transforms according to the action of $\mathfrak h$ on $\mathfrak m$, whereas the principal connection $\omega_\mathfrak h$ transforms by the covariant derivative of the $\mathfrak h$-valued parameter $\lambda_\mathfrak h$.

Finally, suppose that the Cartan geometry is torsion-free,
$$
(F_\omega)_\mathfrak m=0.
$$
Then the variation of the coframe $\omega_\mathfrak m$ induced by $X$ becomes
$$
\pounds_X\omega_\mathfrak m
=
\mathrm{d}_{\omega_\mathfrak h}\lambda_\mathfrak m
+
[\omega_\mathfrak m,\lambda_\mathfrak h]
+
[\lambda_\mathfrak m,\omega_\mathfrak m]_\mathfrak m.
$$

Thus, in the torsion-free case, the infinitesimal variation of the coframe is determined by the covariant derivative of the $\mathfrak m$-valued representative of the vector field induced on the base, together with its $\mathfrak h$-component and, in the non-symmetric case, the $\mathfrak m$-projection of the bracket $[\lambda_\mathfrak m,\omega_\mathfrak m]$.

The main purpose of this paper is to study the Hamiltonian formulation of several gravitational actions. For this reason, we will restrict our considerations to trivial (actually, trivialized) principal bundles.
\[
H \rightarrow P=M\times H \rightarrow M,
\]
equipped with the canonical section
\(s_e:M\to P\),  \(\, s_e(x)=(x,e)\),
where \(e\in H\) is the identity element. By doing this, we can pull back the Cartan connection to a \(\mathfrak g\)-valued 1-form on \(M\),
\[
W=s_e^*\omega\in\Omega^1(M,\mathfrak g).
\]
Another choice of section,
\(s_h(x)=(x,h(x))\),
(with \(h:M\to H\) smooth), induces a gauge transformation
\[
W\mapsto \mathrm{Ad}(h^{-1})\circ W+h^{-1}\ddd h.
\]
For comprehensive discussions of globalization issues and examples see \cite{Freed:1992vw,Tejero:2004}

%
%
	
\section{Actions, symmetries and field equations}\label{sec_Actions}

We start now with the analysis of several actions related to General Relativity. We will not cover all of the existing ones but, rather, look at those that can be written exclusively in terms of differential forms. There are several reasons for this choice. The first is that differential forms, as shown in the preceding section, play a central role in Cartan geometry and have a number of useful geometric features that are not shared, in general, by other types of tensor fields. In particular, we want to highlight the simple interplay between pullbacks and natural operations with differential forms such as the exterior differential and the exterior product. Also, some of the actions that we consider here have some kind of relationship with the Ashtekar formulation for GR. In all the cases, we will show how the used conceptual geometric framework provides solid foundations to understand the details of the Hamiltonian treatment that can be exported to any physical context where Hamiltonians play a relevant role.

\subsection{Geometric preliminaries}\label{subsec_Actions_prelim}

Here we will use the ideas of Cartan geometry summarized in the preceding section to identify the geometric objects that will play a relevant role in the paper and the gauge symmetries of the models that we study.
We start by writing the Lie brackets for $\mathfrak{iso}(1,3)$, $\mathfrak{iso}(4)$ and $\mathfrak{carr}(1,3)$ in a convenient form that will allow us to choose any of them just by selecting the value of a parameter $\zeta=\pm1,0$ [$\zeta=-1$ for $\mathfrak{iso}(1,3)$, $\zeta=+1$ for $\mathfrak{iso}(4)$ and $\zeta=0$ for $\mathfrak{carr}(1,3)$]. In the basis of generators $H\,,P_i\,,J_i\,,K_i$ the Lie brackets are:
\[\begin{blockarray}{llll}
         [H,H]=0 \ \ \mbox{}& [H,K_j]=-\zeta P_j & [H,P_j]=0& [H,J_j]=0   \\
        & [K_i,K_j]=\zeta\varepsilon_{ij}^{\phantom{ij\,}k}K_k   \ \ \mbox{}& [K_i,P_j]=-\delta_{ij} H & [K_i,J_j]=\varepsilon_{ij}^{\phantom{ij\,}k}K_k  \\
        && [P_i,P_j]=0\ \ \mbox{} &[P_i,J_j]=\varepsilon_{ij}^{\phantom{ij\,}k}P_k  \\
          &  &   & [J_i,J_j]=\varepsilon_{ij}^{\phantom{ij\,}k}J_k \\
\end{blockarray}\]
where $i,j,k=1,2,3$, $\varepsilon_{ijk}$ is the 3-dim Levi-Civita symbol and the indices are raised with $\delta^{ij}$. If we denote $\mathfrak{h}=\mathrm{span}\{J_i,K_i\}$ and $\mathfrak{m}=\mathrm{span}\{H,P_i\}$ we have $[\mathfrak{h},\mathfrak{h}]\subset \mathfrak{h}$, $[\mathfrak{h},\mathfrak{m}]\subset \mathfrak{m}$ and $[\mathfrak{m},\mathfrak{m}]=0$, hence, we will be dealing with reductive, symmetric Cartan geometries.

A Lie algebra valued connection 1-form will be expanded in the previous basis as
\[
C=\tau H+e^iP_i+A^iJ_i+\Omega^iJ_i
\]
with $\tau\,,e^i\,,A^i\,,\Omega^i\in\Omega^1(M)$ (here $M$ denotes the spacetime manifold). The curvature of $C$ is
\[
F_C:=\bd C+\frac{1}{2}[C\wed C]\,.
\]
Its expansion in the basis $\{H\,,P_i\,,J_i\,,K_i\}$ is
\[
F_C=(\bd\tau+\langle e\wed \Omega\rangle)H+(\bd_Ae+\zeta\Omega\wedge \tau)^iP_i+\Big(F_A+\frac{\zeta}{2}[\Omega\wed \Omega]\Big)^i J_i+(\bd_A\Omega)^iK_i\,,
\]
where
\begin{equation*}
\begin{array}{l}
\bd_A e:=\bd e+[A\wed e]\,,\\
\bd_A \Omega:=\bd \Omega+[A\wed \Omega]\,,\\
F_A:=\bd A+\frac{1}{2}[A\wed A]\,.
\end{array}
\end{equation*}
The projection of $F_C$ onto $\mathfrak{m}$ is known as the \textit{torsion}
\[
(\bd\tau+\langle e\wed \Omega\rangle)H+(\bd_Ae+\zeta\Omega\wedge \tau)^iP_i\,,
\]
and the projection onto $\mathfrak{h}$ the $H$-\textit{curvature}
\[
\Big(F_A+\frac{\zeta}{2}[\Omega\wed \Omega]\Big)^i J_i+(\bd_A\Omega)^iK_i\,.
\]
In the following, it will be useful to write everything in terms of these objects as they have a clear geometric meaning.

If we take $X_{\mathfrak{h}}=\rho^iJ_i+\beta^iK_i$ with local gauge parameters $\beta_i\,,\rho_i\in\Omega^0(M)$, the gauge transformations of the connection are given by $\delta_{(\rho,\beta)}C:=\bd X_{\mathfrak{h}}+[C,X_{\mathfrak{h}}]$. We have then
\begin{equation}\label{gauge_transformations}
\delta_{(\rho,\beta)}C=\langle\beta\wed e\rangle H+([e\wed \rho]-\zeta\tau\beta)^iP_i+(\bd_A\rho+\zeta[\Omega\wed\beta])^iJ_i+(\bd_A\beta+[\Omega\wed\rho])^iK_i\,.
\end{equation}
These give the gauge transformations of $\tau,e,A,\Omega$. With these ingredients we are ready to write down the actions that we will study here.

\subsection{The Cartan-Palatini, Holst and Carrollian-Holst actions}\label{subsec_Holst_gen}

The main goal of the paper is to use ideas and concepts taken from Cartan geometry and use them to expedite and interpret the Hamiltonian description of GR and Carrollian GR. Let us take a smooth 4-manifold $M=[t_1,t_2]\times\Sigma$ with $\Sigma$ a closed, orientable, and 3-dimensional manifold (hence, parallelizable). Our starting point will be the following action (in fact, a family of actions) parameterized by $\zeta=0\,,\pm1$ and $\gamma\in\mathbb{R}$:

\begin{equation}\label{action_Holst}
\begin{split}
S(\uptau,\ee,\Aa,\Upomega)=\int_{M}L=\int_{M}&\left( \,\langle\Big(\uptau\wedge \ee\,+\frac{1}{2\gamma}[\ee\wed \ee]\Big)\wed\Big(F_A+\frac{\zeta}{2}[\Omega\wed \Omega]\Big)\rangle\right.\\
&\left. +\langle\Big(\frac{\zeta}{\gamma}\uptau\wedge \ee+\frac{1}{2}[\ee\wed \ee]  \Big)\wed\bd_\Aa\Upomega   \rangle\right)
\end{split}
\end{equation}
We have written it in a way that highlights the components of the $\mathfrak{h}$-curvature discussed above. A straightforward computation shows that this action contains several interesting subcases controlled by the values of the parameters $\zeta$ and $\gamma\neq0$:
\begin{align*}
  \zeta & =0\quad&\text{Carrollian Holst}\,,  \\
  \zeta & =-1 \quad&\text{Lorentzian Holst}\,,  \\
  \zeta & =-1\,,\gamma\rightarrow\infty \quad&\text{Lorentzian Cartan-Palatini}\,,\\
  \zeta & =+1 \quad&\text{Euclidean Holst}\,,  \\
  \zeta & =+1\,,\gamma\rightarrow\infty \quad&\text{Euclidean Cartan-Palatini}\,.
\end{align*}
Here Carrollian-Holst gravity refers to the Carroll-invariant action obtained in \cite{Figueroa-OFarrill:2022mcy} from the Holst action for GR. As we will show in the paper, it is possible to study all these actions in a conceptually simple and unified way.

An explicit computation shows that the \eqref{action_Holst} is invariant under infinitesimal transformations of the form given at the end of Section \ref{sec_Cartan_geometry}:
\begin{align*}
\delta_{(\rho,\beta)}\uptau&=\langle\beta\wed \ee\rangle\,, && \delta_{(\rho,\beta)}\ee\,\,=-\zeta\uptau \beta- [\rho\wed \ee]\,,&&\\
\delta_{(\rho,\beta)} \Aa&=\ddd_\Aa\rho-\zeta [\beta\wed \Omega]\,, && \delta_{(\rho,\beta)}\Upomega=\ddd_\Aa\beta-[\rho\wed \Upomega]\,,&&
\end{align*}
with parameters $\rho\in\Omega^0(M)$ for the internal rotations and $\beta\in\Omega^0(M)$ for the boosts---either Lorentz or Carrollian---or the ``other $SU(2)$ factor'' in the Euclidean case (we will take the liberty of referring to all these transformations as ``boosts'' or ``generalized boosts''). The action is also invariant under diffeomorphisms of the spacetime manifold $M$. Before proceeding further, it is important to comment on the way this action is written. First of all, it is obviously invariant under 4-dimensional diffeomorphisms and under (internal) $SU(2)$ transformations. This is so because the inner product $\langle\cdot,\cdot\rangle$ introduced in the Appendix \ref{app_notation} is proportional to the Cartan-Killing form in $SU(2)$. However, the invariance under boosts is hidden in this notation. From a geometric perspective it would be better to write the action in a way that makes all the symmetries apparent, however, as our purpose is to arrive at the Ashtekar formulation, it is more convenient for us to work with \eqref{action_Holst}.

The variation of the Lagrangian that defines the action \eqref{action_Holst} is
\begin{equation}
    \dd L  =\dd\uptau\wedge \mathrm{EOM}_\uptau+\langle\dd\ee\wed\mathrm{EOM}_\ee\rangle+\langle\dd \Aa \wed \mathrm{EOM}_\Aa\rangle+\langle\dd\Upomega\wed \mathrm{EOM}_\Upomega\rangle+\ddd\Theta
\end{equation}
with
\begin{subequations}
\begin{align}
  &\mathrm{EOM}_\uptau:=\langle \ee\wed\big(\FF_\Aa+\frac{\zeta}{2}[\Upomega\wed\Upomega]+\frac{\zeta}{\gamma}\ddd_\Aa\Upomega\big)\rangle\,,\label{Eq_delta_tau}\\
  &\mathrm{EOM}_\ee:=[\ee\wed\Big(\frac{1}{\gamma}\Big(\FF_\Aa+\frac{\zeta}{2}[\Upomega\wed \Upomega]\Big)+\ddd_\Aa\Upomega\Big)]-\uptau\wedge\Big(\FF_\Aa+\frac{\zeta}{2}[\Upomega\wed\Upomega]+\frac{\zeta}{\gamma}\ddd_\Aa\Upomega\Big)\label{Eq_delta_e}\\
  &\mathrm{EOM}_\Aa:= \frac{1}{\gamma}[(\ddd_\Aa\ee-\zeta\uptau\wedge\Upomega)\wed \ee]\,\,+\,\,\ddd_\Aa(\uptau\wedge\ee)-\frac{1}{2}[[\ee\wed\ee]\wed\Upomega]\,,\label{Eq_delta_A}\\
  &\mathrm{EOM}_\Upomega:=\phantom{\frac{1}{\gamma}}\![(\ddd_\Aa\ee-\zeta\uptau\wedge\Upomega)\wed \ee]+\!\frac{\zeta}{\gamma}\Big(\ddd_\Aa(\uptau\wed \ee)-\frac{1}{2}[[\ee\wed\ee]\wed\Upomega]\Big)\,,\label{Eq_delta_Omega}\\
  &\Theta=\langle\Big(\uptau\wedge\ee+\frac{1}{2\gamma}[\ee\wed\ee]\Big)\wed\dd\Aa\rangle+\langle\Big(\frac{\zeta}{\gamma}\uptau\wedge\ee+\frac{1}{2}[\ee\wed\ee]\Big)\wed\dd\Upomega\rangle
\end{align}\end{subequations}
The discussion of Section \ref{sec_Cartan_geometry} suggests that it may be possible to simplify these equations and write them in a more transparent and geometric way. To this end, first notice that if  (\ref{Eq_delta_A},\ref{Eq_delta_Omega}) vanish and $\gamma^2\neq\zeta$, then
\begin{align}
  & 2\bd_\Aa(\uptau\wedge\ee)-[[\ee\wed\ee]\wed\Upomega]=0\,,\label{Field_eq_3_bis}\\
  & [(\ddd_\Aa\ee-\zeta\uptau\wedge\Upomega)\wed \ee]=0\label{Field_eq_2_bis}\,,
\end{align}
($\gamma^2=\zeta$ corresponds to the self-dual case, which we do not consider in this paper). It is now possible (see appendix \ref{sub_app_simpli}) to show that these equations are equivalent to
\begin{align}
  & \ddd\uptau+\langle\ee\wed\Upomega\rangle=0\label{Field_eq_3_ter}\,,\\
  & \ddd_\Aa\ee+\zeta\Upomega\wedge\uptau=0\label{Field_eq_2_ter}\,,
\end{align}
which can be immediately interpreted as vanishing torsion conditions. The remaining equations (\ref{Eq_delta_e},\ref{Eq_delta_tau}) set equal to zero can be written in the form
\begin{align}
&\langle\ee\wed \Big(\FF_\Aa+\frac{\zeta}{2}[\Upomega\wed\Upomega]\Big)\rangle=0\,,\label{Field_eq_1_bis}\\
&\uptau\wedge\Big(\FF_\Aa+\frac{\zeta}{2}[\Upomega\wed\Upomega]\Big)-[\ee\wed\ddd_\Aa\Upomega]=0\,.\label{Field_eq_4_bis}
\end{align}
by using the auxiliary conditions
\begin{align}
& \langle\ee\wed\bd_\Aa\Upomega\rangle=0\,,\label{c1}\\
&[\ee\wed\Big(\FF_\Aa+\frac{\zeta}{2}[\Upomega\wed\Upomega]\Big)]-\zeta\uptau\wedge\bd_\Aa\Upomega=0\,,\label{c2}
\end{align}
obtained  by  taking the covariant exterior derivative $\ddd_\Aa$ of (\ref{Field_eq_3_ter},\ref{Field_eq_2_ter}). As we can see, irrespective of the value of $\zeta$, the Immirzi parameter $\gamma$ drops from the final form of the field equations (\ref{Field_eq_3_ter},\ref{Field_eq_2_ter},\ref{Field_eq_1_bis},\ref{Field_eq_4_bis}).

We can also obtain the symplectic current \cite{Margalef-Bentabol:2020teu,Margalef-Bentabol:2022zso} by computing $-\dd\Theta$ which, after some straightforward manipulations can be written as:
\begin{equation}\label{eq: cov symplectic current}
    -\dd\Theta=\langle\dd\Aa\wed\dd\Big(\frac{1}{2\gamma}[\ee\wed\ee]-\ee\wedge\uptau\Big)\rangle+\langle\dd\Upomega\wed\dd\Big(\frac{1}{2}[\ee\wed\ee]-\frac{\zeta}{\gamma}\ee\wedge\uptau\Big)\rangle
\end{equation}

In order to disentangle the meaning of the dynamics described by the field equations (\ref{Field_eq_3_ter},\ref{Field_eq_2_ter},\ref{Field_eq_1_bis},\ref{Field_eq_4_bis}), it is useful to introduce the frame dual to the co-tetrad $(\uptau,\ee)$ which we write in the form $(\uu,\E)$ and satisfies
\[
\uu\iprod\uptau=1\,,\E\iprod\uptau=0\,,\uu\iprod \ee=0\,,\E\iprod\ee=\mathbbm{1}\,.
\]
By making use of the field equations, the Lie derivatives of $\{\uptau\,,\ee\,,\Aa\,,\Upomega\}$ along $\uu$ can be written as
\begin{align*}
&\pounds_\uu\uptau=\langle\ee\wed(\uu\iprod\Upomega)\rangle\,,\\
&\pounds_\uu\ee=-\zeta(\uu\iprod\Upomega)\uptau-[(\uu\iprod\Aa)\wed\ee]+\zeta\Upomega\,,\\
&\pounds_\uu\Aa=\bd_\Aa(\uu\iprod\Aa)-\zeta[(\uu\iprod\Upomega)\wed\Upomega]+\zeta\ast_\ee(\underline{\bd_\Aa\Upomega})\,,\\
&\pounds_\uu\Upomega=\bd_\Aa(\uu\iprod\Upomega)-[(\uu\iprod\Aa)\wed\Upomega+\ast_\ee\Big(\underline{\FF_\Aa}+\frac{\zeta}{2}[\underline{\Upomega}\wed \underline{\Upomega}]\Big)\,.
\end{align*}
The notation used in the previous expressions is explained in (\ref{subsec_Lie}), some calculational details are also given there. As we can see, these Lie derivatives contain a gauge part corresponding to internal rotations with parameter $\uu\iprod\Aa$ and generalized boosts with parameter $\uu\iprod\Upomega$. Except in the case of $\pounds_\uu\uptau$, there are also nontrivial contributions with rather simple expressions. Notice that in the Carrollian case ($\zeta=0$) the only dynamics that is not pure gauge is that of $\Upomega$.

We look now at the symmetric metric tensor
\[
\Gg:=\zeta\uptau\otimes\uptau+\langle\ee\otimes\ee\rangle\,.
\]
This object can be seen to be invariant under rotations and generalized boosts. For $\zeta=-1$ it is a Lorentzian metric, a degenerate metric in the Carrollian case $\zeta=0$, and a Euclidean metric for $\zeta=+1$. Its Lie derivative along $\uu$ is
\begin{equation}\label{Lie_g}
\pounds_\uu\Gg=\zeta\big(\langle(\uu\iprod\Upomega)\wed\ee\rangle\otimes\uptau+\uptau\otimes\langle(\uu\iprod\Upomega)\wed\ee\rangle
+\langle\underline{\Upomega}\otimes\ee\rangle+\langle\ee \otimes\underline{\Upomega}\rangle  \big)\,,
\end{equation}
where we have expanded $\Upomega^i=\Omega^{ij}\ee_j+\eta^i\uptau=:\underline{\Upomega}+\eta^i\uptau\,.$ Notice that \eqref{Lie_g} vanishes in the Carrollian case $\zeta=0$.

It is interesting to study what happens when $\uptau$ defines an integrable distribution. According to Frobenius theorem, this happens when $\uptau\wedge\bd\uptau=0$. Now, using the field equation \eqref{Field_eq_3_ter} and the expansion for $\Upomega$ in terms of $\Omega^{ij}$ and $\eta^i$ this integrability condition becomes
\[
\varepsilon_{ijk}\Omega^{jk}=0\Leftrightarrow \langle\ee\wed\underline{\Upomega}\rangle=0\,.
\]
This has the following interesting consequence: the integral submanifolds of the distribution defined by $\uptau$ have extrinsic curvature given by
\[
K_ij\ee^i\otimes\ee^j=\frac{1}{2}\pounds_\uu\langle\ee\otimes\ee\rangle=\zeta\underline{\Upomega}_{ij}\ee^i\otimes\ee^j\Rightarrow K_{ij}=\zeta\underline{\Upomega}_{ij}\,.
\]
Thus, in the Carrollian case, the extrinsic curvature is always zero irrespective of $\Omega$.

We end this section with the following comment: for the kind of spacetime manifolds and field configurations that we use in the paper, any given $\uptau$ is gauge related to an integrable one $\tilde{\uptau}$. This is discussed at the end of section \ref{sec_Hamiltonians}.

%
%
\section{Hamiltonian formulation}\label{sec_Hamiltonians}



In this section we provide a concise and conceptually simple derivation of the Hamiltonian formulation for the action \eqref{action_Holst}. As mentioned in the introduction, we will take advantage of the fact that the field equations are linear in first order derivatives of the fields to arrive at the Hamiltonian formulation in a way that avoids the intricacies of other methods such as the Dirac approach \cite{Dirac2} or GNH \cite{Gotay1,Gotay2,Margalef-Bentabol:2018}. As discussed in Section \ref{sec_DiracGNH}, the result is the same. There are several observations that justify this. On one hand, the fact that the action only involves first order derivatives means that the computation of the canonical momenta leads immediately to primary constrains which do not involve any velocities and are, thus, just functions of the configuration variables. This means that the primary constraint submanifold in phase space will be parametrized by the configuration variables. In addition to this, as the action is homogeneous of first degree in velocities, the energy will be just a function of the configuration variables. This implies that the Hamiltonian will be equal to the energy and will be well-defined on the primary constraint submanifold  $\mathsf{M}_0$. If one uses the GNH approach to find the Hamiltonian description of the dynamics, the Hamiltonian vector fields, tangent to $\mathsf{M}_0$ will just give us the time derivatives of the configuration variables. As we show next, these can be easily found directly from the field equations obtained in the previous section by a simple projection procedure.

The spacetime manifold introduced in Section \ref{sec_Actions} is foliated by ``spatial slices'' of the form $\Sigma_t:=\{t\}\times\Sigma$ ($t\in[t_1,t_2]$). Since $M$ is a Cartesian product, we have some canonically defined objects that we will extensively use in the following. On one hand, we have a evolution vector field $\partial_{\mathrm{t}}\in \mathfrak{X}(M)$, transverse to every $\Sigma_t$ which is defined by the tangent vectors to the curves $c_p:\mathbb{R}\rightarrow M=[t_1,t_2]\times \Sigma:t\mapsto (t,p)$. We also have a scalar field $t\in C^\infty(M)$ given by $t:M\rightarrow \mathbb{R}:(\tau,p)\mapsto \tau$, such that ${\partial_{\mathrm{t}}}\iprod \bd t=1$. For each $t\in[t_1,t_2]$, we define the embedding $\jmath_t:\Sigma\rightarrow M:p\mapsto(t,p)$ with its pullback denoted as $\jmath_t^\ast$. Notice that $\Sigma_t=\jmath_t(\Sigma)$.

If we pullback the field equations (\ref{Field_eq_3_ter},\ref{Field_eq_2_ter},\ref{Field_eq_1_bis},\ref{Field_eq_4_bis}) onto each $\Sigma_t$ ($t\in[t_1,t_2]$), we find the following conditions
\begin{equation}\label{eq: constraints}
\begin{aligned}
&\mathcal{C}_1:=\ddd\tau-\langle\Omega\wed e\rangle=0\,, &\mathcal{C}_2:=\ddd_A e-\zeta\tau\wedge\Omega=0\,, \hspace*{35mm}\\
&\mathcal{C}_3:=\langle e\wed \Big(F_\Aa+\frac{\zeta}{2}[\Omega\wed\Omega]\Big)\rangle=0\,,\,\,& \mathcal{C}_4:=\tau\wedge\Big(F_\Aa+\frac{\zeta}{2}[\Omega\wed\Omega]\Big)-[e\wed\ddd_A\Omega]=0\,,
\end{aligned}
\end{equation}
where $\displaystyle\tau(t):=\jmath_t^*\uptau\,,e(t):=\jmath_t^*\ee\,,A(t):=\jmath_t^*\Aa\,,\Omega(t):=\jmath_t^*\Upomega\,.$ These objects depend on the time parameter $t$  (although we have written $\tau$ instead of $\tau(t)$, and so on, to lighten the notation). We have also used the identities
\begin{align*}
&\jmath_{\mathrm{t}}^*\FF_\Aa=\ddd A+\frac{1}{2}[A\wed A]=:F_A&&\jmath_{\mathrm{t}}^*\ddd_\Aa\uplambda=\ddd\uplambda+[A\wed\uplambda]=:\ddd_A\lambda\,.
\end{align*}
The conditions \eqref{eq: constraints} are constraints that must hold for every $t\in[t_1,t_2]$.

We can also contract the field equations with the vector field $\partial_{\mathrm{t}}$ and then pullback the result onto $\Sigma_t$. By doing this, we get the following evolution equations:
\begin{subequations}
\begin{align}
    &0=\dot{\tau}-\ddd\taut-\langle\Ot\wed e\rangle+\langle\Omega\wed\et\rangle\,.\label{eq: dynamic4}\\
    &0=\dot{e}-\ddd_A\et+[\At\wed e]-\zeta \taut\Omega+\zeta\Ot\tau\,,\label{eq: dynamic3}\\
    &0= \langle\et\wed\Big(F_\Aa+\frac{\zeta}{2}[\Omega\wed\Omega]\Big)\rangle-\langle e\wed X_A\rangle \label{eq: dynamic1}\,, \\
    &0=[\et\wed\ddd_A\Omega]-[e\wed X_\Omega]-\taut \Big(F_\Aa+\frac{\zeta}{2}[\Omega\wed\Omega]\Big)+\tau\wedge X_A\label{eq: dynamic2}\,,\\
    &\dot{\taut},\dot{e}_{\mathrm{t}},\dot{\At},\dot{\Omega}_{\mathrm{t}}, \text{ arbitrary,}
\end{align}\end{subequations}
where
\begin{align*}
 & \tau_{\mathrm{t}}:=\jmath_t^*\partial_{\mathrm{t}}\iprod\uptau\,, &&e_{\mathrm{t}}:=\jmath_t^*\partial_{\mathrm{t}}\iprod\ee\,, &&A_{\mathrm{t}}:=\jmath_t^*\partial_{\mathrm{t}}\iprod\Aa\,, &&\Omega_{\mathrm{t}}:=\jmath_t^*\partial_{\mathrm{t}}\iprod\Upomega\,, \\
 & \,\dot{\tau}:=\jmath_t^*\pounds_{\partial_{\mathrm{t}}}\uptau\,,&&\dot{e}:=\jmath_t^*\pounds_{\partial_{\mathrm{t}}}\ee\,,&&\dot{A}:=\jmath_t^*\pounds_{\partial_{\mathrm{t}}}\Aa\,,&&\dot{\Omega}:=\jmath_t^*\pounds_{\partial_{\mathrm{t}}}\Upomega\,, \end{align*}
and we use the short-hand notation
\[X_A:=\dot{A}-\ddd_A\At+\zeta[\Ot\wed\Omega]\,,\qquad \qquad X_\Omega:=\dot{\Omega}-\ddd_A\Ot+[\At\wed\Omega]\,.\]
We have also used the identities:
\begin{align*}
&\jmath_t^*(\partial_t\iprod\FF_\Aa)=\dot{A}-\ddd_A \At\,,&& \jmath_t^*(\partial_t\iprod\ddd_\Aa\uplambda)=\dot{\lambda}-\ddd_A \lambda_{\mathrm{t}} +[\At\wed\lambda]
\end{align*}
Notice that \eqref{eq: dynamic4} and \eqref{eq: dynamic3} provide directly the dynamics for $\tau$ and $e$, whereas, for $A$ and $\Omega$, we need to solve \eqref{eq: dynamic1} and \eqref{eq: dynamic2} for $\dot{A}$ and $\dot{\Omega}$. To this end, it is useful to introduce the following objects
\begin{equation}\label{eq: gauge parameters}
    \xi\iprod e:=\et\,,\qquad \Lambda:=\At-\xi\iprod A\,,\qquad N:=\taut-\xi\iprod\tau\,,\qquad\Psi:=\Ot-\xi\iprod\Omega\,,
\end{equation}
which allow us, using the constraints \eqref{eq: constraints}, to rewrite equations \eqref{eq: dynamic4}-\eqref{eq: dynamic2} as:
\begin{align}
    &\dot{\tau}=\pounds_\xi\tau+\langle\Psi\wed e\rangle+\ddd N\,.\label{eq: rewritten dynamic4}\\
    &\dot{e}=\pounds_\xi e-[\Lambda\wed e]+\zeta(N\Omega-\Psi\tau)\,,\nonumber\\
    &\dot{A}=\pounds_\xi A+\ddd_A\Lambda-\zeta[\Psi\wed\Omega]+Z_A, \text{with }\mathrm{E}_1:=\langle e\wed Z_A\rangle=0\,,\nonumber\\
    &\dot{\Omega}=\pounds_\xi\Omega+\ddd_A\Psi-[\Lambda\wed\Omega]+Z_\Omega,\hspace*{1mm} \text{ with }\mathrm{E}_2:=[ e\wed Z_\Omega]-\tau\wedge Z_A+N\Big(F_\Aa+\frac{\zeta}{2}[\Omega\wed\Omega]\Big)=0\,.\nonumber
\end{align}
Notice that the vector field $\xi$ is implicitly defined in terms of $e$ and $\et$ by the condition $\xi\iprod e=\et$.

We still have to solve for $Z_A$ and $Z_\Omega$ and check the consistency of the constraints under time evolution. In order to do this we compute the derivatives of the constraints with respect to $t$. We find
\begin{equation*}
\dot{\mathcal{C}_1}=\pounds_\xi\mathcal{C}_1+\langle\mathcal{C}_2\wed\Psi\rangle+\overbrace{\langle e\wed  Z_\Omega\rangle}^{\mathrm{E}_3}\stackrel{.}{=}\mathrm{E}_3\hspace*{6.2cm}
\end{equation*}

\vspace*{-11mm}

\begin{align*}
\dot{\mathcal{C}_2}&=\pounds_\xi\mathcal{C}_2+[\mathcal{C}_2 \wed\Lambda]-\zeta\Psi \mathcal{C}_1+\overbrace{[e\wed Z_A]-\zeta\tau\wedge Z_\Omega+\zeta N\ddd_A\Omega}^{\mathrm{E}_4}\stackrel{.}{=}\mathrm{E}_4\\
\dot{\mathcal{C}_3}&=\pounds_\xi\mathcal{C}_3-\zeta\langle\mathcal{C}_4\wed\Psi\rangle+\langle\mathcal{C}_2\wed Z_A\rangle+\zeta\langle\Omega\wed \mathrm{E}_2\rangle-\ddd\mathrm{E}_1\stackrel{.}{=}0\\
\dot{\mathcal{C}_4}&=\pounds_\xi\mathcal{C}_4+[ \mathcal{C}_4\wed\Lambda]+\mathcal{C}_3\wedge\Psi+\mathcal{C}_1\wedge Z_A-[\mathcal{C}_2\wed Z_\Omega]+\mathrm{E}_1\wedge\Omega+\ddd_A\mathrm{E}_2\stackrel{.}{=}0\,,
\end{align*}
where $\stackrel{.}{=}$ signals that we have used the constraints. As we can see, the last two conditions hold automatically if $\mathrm{E}_1=0$, $\mathrm{E}_2=0$ (and using the constraints), hence, to complete the Hamiltonian analysis it remains to solve $\{\mathrm{E}_I=0\}_{I=1,2,3,4}$ for $(Z_A,Z_\Omega)$ that we write as
\begin{align}
  & \langle  Z_A\wed e\rangle=0\,,\label{EE1}\\
  &Z_A\wedge\tau+[Z_\Omega\wed e]=-N F_{(A,\Omega)}\,, \label{EE2}\\
  & \langle Z_\Omega\wed e \rangle=0\,,\label{EE3}\\
  &[Z_A\wed e]+\zeta Z_\Omega\wedge \tau =-\zeta N D\,, \label{EE4}
\end{align}
To showcase the ``symmetry'' of the equations, we have introduced the shorthand notation
\begin{align*}
&D:=\ddd_A\Omega\,,\\
&F_{(A,\Omega)}:=F_A+\frac{\zeta}{2}[\Omega\wed\Omega]\,.
\end{align*}
In order to solve these equations we expand
\begin{equation*}
\begin{array}{ll}
Z_A^i=a^i_{\phantom{i}j}e^j\,, &D^i=\frac{1}{2}\mathsf{D}^i_{\phantom{i}j}\varepsilon^{j}_{\phantom{j}k\ell}e^k\wedge e^\ell\,,\phantom{\Big(}\\
Z_\Omega^i=\omega^i_{\phantom{i}j}e^j\,, &F_{(A,\Omega)}^i=\frac{1}{2}\mathsf{P}^i_{\phantom{i}j}\varepsilon^{j}_{\phantom{j}k\ell}e^k\wedge e^\ell\,,\phantom{\bigg(}\\
\tau=\tau_i e^i\,.&
\end{array}
\end{equation*}
Before proceeding further, it is useful to write down the conditions that the constraints (\ref{eq: constraints}) imply on $\mathsf{D}_{ij}$ and $\mathsf{P}_{ij}$. They are, respectively
\begin{align}\label{Rels}
\begin{split}
& \mathsf{D}^i_{\phantom{i}i}=0\,,\\
& \zeta\mathsf{D}_{ij}\tau^j+\varepsilon_{ijk}\mathsf{P}^{jk}=0\,, \\
& \mathsf{P}^i_{\phantom{i}i}=0\,, \\
& \mathsf{P}_{ij}\tau^j+\varepsilon_{ijk}\mathsf{D}^{jk}=0\,.
\end{split}
\end{align}
The first two come from the covariant exterior differential $\ddd_A$ of $\mathcal{C}_1$ and $\mathcal{C}_2$, whereas the last two come from $\mathcal{C}_3$ and $\mathcal{C}_4$.

Plugging now the expansions given above into (\ref{EE1}-\ref{EE4}) gives the following equations for $a_{ij}$ and $\omega_{ij}$
\begin{align*}
a^k_{\phantom{k}k}\delta^{ij}-a^{ij}+\zeta \tau_k \omega^j_{\phantom{i}\ell}\varepsilon^{\ell k i}=\-\zeta N \mathsf{D}^{ji}\,,&\qquad \varepsilon_{ijk}a^{jk}=0\,,\\
\omega^k_{\phantom{k}k}\delta^{ij}-\,\omega^{ij}\,+\,\tau_k a^j_{\phantom{i}\ell}\varepsilon^{\ell k i}=\-\zeta N \mathsf{P}^{ji}\,,&\qquad \varepsilon_{ijk}\omega^{jk}=0\,,
\end{align*}
Their solutions are
\begin{align*}
&a_{ij}=\,\frac{\zeta N}{1+\zeta \tau^2}\big(\mathsf{D}_{ij}+\mathsf{P}_i^{\phantom{i}k}\tau^\ell\varepsilon_{k\ell j}-\varepsilon_{ik\ell}\mathsf{P}^{k\ell}\tau_j\big)\,,\\
&\omega_{ij}=\frac{N}{1+\zeta \tau^2}\big(\mathsf{P}_{ij}+\zeta\mathsf{D}_i^{\phantom{i}k}\tau^\ell\varepsilon_{k\ell j}-\zeta\varepsilon_{ik\ell}\mathsf{D}^{k\ell}\tau_j\big)\,.
\end{align*}
where $\tau^2:=\tau^i\tau_i$. By using the Hodge dual as explained in \ref{subsec_Hodge} and \ref{subsec_duals}, we arrive at the final form for the components of the Hamiltonian vector fields:
\begin{align}
&X_\tau\,=\pounds_\xi\tau+\ddd N+\langle\psi\wed e\rangle\,, \phantom{\bigg)}\nonumber\\
&X_e\,=\pounds_\xi e-\zeta \tau\psi+[e\wed\Lambda]+\zeta N \Omega\,,\nonumber\\
&X_A=\pounds_\xi A+\ddd_A\Lambda+\zeta[\Omega\wed\psi]+\frac{\zeta N}{1+\zeta \tau^2}\big(\!*\!\{\bd_A\Omega-\tau\wedge *F_{(A,\Omega)}\}+*(\langle *F_{(A,\Omega)}\wed e\rangle\wedge e)\,\tau \big)\,,\nonumber\\
&X_\Omega=\pounds_\xi \Omega+\ddd_A\psi+[\Omega\wed\Lambda]+\frac{N}{1+\zeta \tau^2}\big(\!*\!\{F_{(A,\Omega)}-\zeta\!(\tau\wedge *\bd_A\Omega)\}+\zeta*\!(\langle *\bd_A\Omega\wed e\rangle\wedge e)\,\tau\big)\,,\nonumber\\
&X_{\taut},X_{\et}, X_{\At},X_{\Ot} \text{ arbitrary}\,,\label{final_vector fields}
\end{align}

\noindent with $\Lambda$, $\psi$ and $N$ given by (\ref{eq: gauge parameters}).

Finally, the symplectic form can be easily obtained by pulling back \eqref{eq: cov symplectic current} to $\Sigma$ and integrating:
\begin{equation}\label{presymplectic_Omega}
        \omega_1=\int_\Sigma \left(\langle\dd A\wwed\dd\Big(\frac{1}{2\gamma}[e\wed e]+(\tau\wedge e)\Big)\rangle+\langle\dd \Omega\wwed\dd\Big(\frac{1}{2}[e\wed e]+\frac{\zeta}{\gamma}(\tau\wedge e)\Big)\rangle\right)\,,
\end{equation}
which is clearly degenerate since the dynamic variables $\et$, $\At$, $\Ot$ and $\taut$ are absent.

From a classical point of view, this Hamiltonian description is completely satisfactory. If initial data satisfying the constraints \eqref{eq: constraints} are given, the integral curves given by \eqref{final_vector fields} define the evolution of those initial data. There is a caveat, though: the dynamical variables $\et$, $\At$, $\Ot$ and $\taut$ are arbitrary at this stage, hence, the evolution is not uniquely defined. This can be solved by gauge fixing: if we choose  $\hat{\xi}\in\mathfrak{X}(\Sigma)$, $\hat{\At}\in\Omega^0(\Sigma,\mathfrak{su}(2))$, $\hat{\Ot}\in\Omega^0(\Sigma,\mathfrak{su}(2))$, $\hat{\taut}\in\Omega^0(\Sigma)$ and fix $\xi=\hat{\xi}$, $\At=\hat{\At}$, $\Ot=\hat{\Ot}$ and $\taut=\hat{\taut}$ the arbitrariness is eliminated. Notice that $\et$ is not fixed. Rather, it is now related to $e$ via $\et=\hat{\xi}\iprod e$. Also, as we want to build non-degenerate triads in 4 dimensions, $\taut$ should be chosen in such a way that the initial data define a non-degenerate tetrad. We assume these conditions hold from now on. Despite this, notice that $\omega_1$ is still degenerate.

We end this section by wrapping up the comment at the end of section \ref{sec_Actions}. For the type of spacetime manifolds that we consider here, and keeping in mind that we choose configurations such that the pullbacks of the $\ee$ onto $\Sigma_t$ are nondegenerate, it is easy to prove that any given 1-form field $\uptau$ is gauge related to one defining an integral distribution. If we use coordinates $(t,x^i)$ in $M$ such that the integral curves of $\partial_t$ are given by $x^i=$constant we can expand
\begin{align*}
\uptau=\tau_0\bd t+\sigma_i\bd x^i\,,\\
\ee^i=e^i_0\bd t+e^i_{\phantom{i}j}\bd x^i\,,
\end{align*}
with $e^i_{\phantom{i}j}$ invertible. Now
\[
\uptau+\langle\beta\wed \ee\rangle=(\tau_0+\beta_ie^i_0)\bd t+(\sigma_i+\beta_ie^i_{\phantom{i}j})\bd x^j\,,
\]
hence by solving $\sigma_i+\beta_ie^i_{\phantom{i}j}=0$ for $\beta_i$ we have the desired integrable $\tilde{\uptau}$ given by $\tilde{\uptau}=(\tau_0+\beta_ie^i_0)\bd t$.

\section{Time gauge}\label{sec_time_gauge}

For the purposes of canonical quantization, the ``non-canonical'' form of $\omega_1$ is problematic. This can be solved by finding appropriate variables on $\mathsf{M}_0$. The best way to do this is by considering an additional gauge fixing, namely, the time gauge. This procedure is discussed in many references (see, for instance, \cite{BarroseSa:2000vx,Alexandrov:2000jw }). Here we closely follow \cite{BarberoG:2020tit}. The key step is to impose the condition $\tau=0$, which in turn implies $\dot{\tau}=0$:
\begin{equation}\label{condition_psi}
0=\dot{\tau}\overset{\eqref{eq: rewritten dynamic4}}{=}\langle\Psi\wed e\rangle+\ddd N\overset{\eqref{eq: gauge parameters}}{=}\ddd\taut+\langle(\Ot-\xi\iprod\Omega)\wed e\rangle=\ddd\taut+\langle\Psi\wed e\rangle\,,
\end{equation}
which can be solved for $\Psi$ in terms of $\taut$, $\et$, $\Omega$ and $e$.

Plugging $\tau=0$ into $\mathcal{C}_2=0$, leads to a unique solution for $A$ in terms of $e$, namely, $A=\Gamma$, where $\Gamma$ is the spin connection (Levi-Civita) satisfying $\ddd_\Gamma e=0$. Taking this into account, the constraints \eqref{eq: constraints} become:
\begin{equation}\label{eq: constraints time gauge}\begin{aligned}
& \overline{\mathcal{C}}_1:=\langle\Omega\wed e\rangle=0\,, &&\qquad\qquad \overline{\mathcal{C}}_2:=A-\Gamma=0\,,\\
& \overline{\mathcal{C}}_3:=\langle e\wed \Big(F_\Gamma+\frac{\zeta}{2}[\Omega\wed \Omega]\Big)\rangle=0\,,&&\qquad\qquad \overline{\mathcal{C}}_4:=-[e\wed\ddd_\Gamma\Omega]=\ddd_\Gamma[e\wed\Omega]=0\,.
\end{aligned}\end{equation}
The evolution of the basic fields is
\begin{subequations}\label{eq: final componens}
\begin{align}
&\dot{e}=\pounds_\xi e-[\Lambda\wed e]+\zeta\tau_t\Omega \,,\label{eq: rerewritten dynamic3}\\
&\dot{\Omega}=\pounds_\xi\Omega+\ddd_\Gamma\Psi-[\Lambda\wed\Omega]+\tau_t\!\ast\!\Big(F_\Gamma+\frac{\zeta}{2}[\Omega\wed \Omega]\Big)\label{eq: rerewritten dynamic2}\,,\\
&\dot{\At},\dot{e}_{\mathrm{t}},\dot{\taut}\,\mathrm{arbitrary}\label{eq: rerewritten dynamic4}
\end{align}\end{subequations}
It is possible to show that, as it should, $\dot{\Gamma}=\dot{A}$ in agreement with \eqref{final_vector fields}:
\[
\dot{\Gamma}=\pounds_\xi \Gamma+\ddd_\Gamma\Lambda-\zeta[\Psi\wed\Omega]+\zeta \tau_t\!\ast\!\ddd_\Gamma\Omega\,.
\]
Finally, the presymplectic form \eqref{presymplectic_Omega} becomes
\begin{equation}\label{Omega_time_gauge}
\overline{\omega}_1=\int_\Sigma\langle\dd(\Gamma+\gamma \Omega)\wwed\dd\Big(\frac{1}{2\gamma}[e\wed e]\Big)\rangle\,,
\end{equation}

The form of the (pre)symplectic form suggests the introduction of the Ashtekar connection
\begin{equation}\label{Ashtekar_conn}
\mathcal{A}:=\Gamma+\gamma\Omega\,.
\end{equation}
and the $\mathfrak{su}(2)$-valued 2-form
\begin{equation}\label{E}
E:=\frac{1}{2}[e\wed e]\,.
\end{equation}
which would render $\mathcal{A}$ and $\g E$ ``canonically conjugate''. The problem with these variables is that the form of the constraints written in terms of them is quite involved because it is difficult to write $e$ directly in terms of $E$. The way out of this predicament is to choose a fiducial, non-dynamical volume form $\mathsf{vol}_0$ (remember that $\Sigma$ is orientable) and introduce a vector field $\widetilde{E}$ defined by
\[
\gamma \widetilde{E}\iprod \mathsf{vol}_0=E\,.
\]
In terms of these variables we can write
\begin{equation}\label{Omega_Asht_2}
\overline{\omega}_1=\int_\Sigma\langle\dd\mathcal{A}\veee\dd \widetilde{E}\rangle\mathsf{vol}_0\,,
\end{equation}
an expression that should be understood as meaning
\[
\overline{\omega}_1(\mathbb{X},\mathbb{Y})=\int_\Sigma\Big(\langle Y_{\widetilde{E}}\iproddot X_{\mathcal{A}} \rangle-\langle X_{\widetilde{E}}\iproddot Y_{\mathcal{A}}\rangle\Big)\mathsf{vol}_0
\]
for $\mathbb{X}=(X_{\mathcal{A}},X_{\widetilde{E}})$ with $X_{\mathcal{A}}$ an $\mathfrak{su}(2)$-valued 1-form on $\Sigma$ and $X_{\widetilde{E}}$ an $\mathfrak{su}(2)$-valued vector field on $\Sigma$. This is the usual sense in which the variables $\mathcal{A}$ and $\widetilde{E}$ are declared to be canonically conjugate. In terms of them the constraints have the form (see \cite{BarberoG:2020tit} for calculational details)
\begin{align}
  & \mathrm{div}_0\widetilde{E}-[\widetilde{E}\iproddot \mathcal{A}]=0\,, \label{Gauss_law}\\
  & \langle \widetilde{E}\iproddot F_{\mathcal{A}}\rangle=0\,, \label{vector_constr}\\
  & \langle \widetilde{E}\iproddot [ \widetilde{E}\iproddot  \big( (\gamma-\frac{\zeta}{\gamma})F_\Gamma+\frac{\zeta}{\gamma} F_{\mathcal{A}} \big)]   \rangle=0\,,\label{scalar_const}
\end{align}
where the divergence $\mathrm{div}_0$ of a vector field $X\in\mathfrak{X}(\Sigma)$ associated with the volume form $\mathsf{vol}_0$ is defined by
\[
\mathrm{div}_0X:=\left(\frac{\pounds_X\mathsf{vol}_0}{\mathsf{vol}_0}\right)\,,
\]
and $\iproddot$ is the obvious generalization of the interior product to include Lie algebra brackets.

To end, we give the form of the Hamiltonian vector fields that give the evolution of the new variables:
\begin{align}\label{components_Ahstekar}
\begin{split}
&\dot{\mathcal{A}}=\pounds_{\xi}\mathcal{A}+\bd_{\mathcal{A}}\Big(\Lambda+\frac{\zeta}{\gamma}\Psi\Big)+\Big(\gamma-\frac{\zeta}{\gamma}\Big)\bd_\Gamma\Psi+
\tau_{\mathrm{t}}\ast\left(\frac{\zeta}{\gamma}F_{\mathcal{A}}+\Big(\gamma-\frac{\zeta}{\gamma}\Big)F_\Gamma\right)\,,\\
&\dot{\widetilde{E}}=\pounds_{\xi}\widetilde{E}+(\mathsf{div}_0\xi)\widetilde{E}-[\Big(\Lambda+\frac{\zeta}{\gamma}\Psi\Big),\widetilde{E}]+
\frac{\zeta}{\gamma}\widetilde{Y}\,,
\end{split}
\end{align}
with $e$ written in terms of $\widetilde{E}$ as shown in \ref{subsec_topforms} and $\widetilde{Y}$ the solution to
\[
\gamma\widetilde{Y}\iprod\mathsf{vol}_0=\bd_{\mathcal{A}}(\tau_{\mathrm{t}}e)\,.
\]
The form of these expressions is very suggestive. On one hand we have terms with the form of Lie derivatives and internal $SU(2)$ rotations defined by the field-dependent objects $\xi$, $\Lambda$ and $\Psi$. Notice that, at this stage there is still some arbitrariness in the dynamics because $\At,e_{\mathrm{t}},\taut$ (and, hence, $\xi,\Lambda, \tau_{\mathrm{t}}$) are still undetermined. This can be solved by choosing appropriate values for these objects (actually, \textit{gauge fixing} them). For instance, we can take $\xi$ equal to a particular vector field $\hat{\xi}\in\mathfrak{X}(\Sigma)$ and fix the values of $\Lambda$ and $\tau_{\mathrm{t}}\neq 0$. When this is done the expressions for the Lie derivatives and the internal rotations become very simple. In fact, nothing precludes us from choosing $\hat{\xi}=0$ and $\gamma\lambda+\zeta\Psi=0$ and remove this part of the dynamics. However, $\tau_{\mathrm{t}}$ must be different from zero because it plays the role of the lapse ($\hat{\xi}$ is essentially the shift). In fact, the $\tau_{\mathrm{t}}$ dependent terms that appear at the end of the expressions for $\dot{\mathcal{A}}$ and $\dot{\widetilde{E}}$ give the non trivial dynamics of the different actions that we consider in the paper.

It is straightforward to check that the evolution given by \ref{components_Ahstekar} corresponds to the one obtained by computing the Poisson brackets of the constraints with $\mathcal{A}$ and $\widetilde{E}$ if they are weighted with the appropriate objects.

%
%
\section{Some comments on the Dirac and GNH methods}{\label{sec_DiracGNH}}

As we have shown in section \ref{sec_Hamiltonians} there is a clean and direct way to obtain the Hamiltonian formulation for each member of the parametric family of actions that we have considered in the paper. An obvious issue that one would like to understand at this point is how does this formulation compares with the ones derived by applying the GNH or Dirac methods and also how the geometric ideas used to arrive at it can illuminate the effective use of these approaches (see \cite{BarberoG:2026urv} for the GNH Hamiltonian study of the Carrollian case $\zeta=0$).

Let us start with the GNH approach. An advantage of the GNH method over Dirac's algorithm is that it is often possible to avoid computing Poisson brackets and, instead, perform computations by comparing the action of relevant differential forms (the exterior differential of the Hamiltonian or objects related to the symplectic form) on vector fields. This is especially useful when dealing with field theories defined on manifolds with boundary \cite{Barbero:2021aul}. In the GNH framework, the starting point is a Lagrangian defined on the tangent bundle $TQ$ of some appropriate configuration space $Q$. In order to derive the corresponding Hamiltonian formulation one has to compute the fiber derivative $FL:TQ\rightarrow T^*Q$ which provides the ``definition of momenta'' in the physics parlance. If this map is a local diffeomorphism the Lagrangian is called \textit{regular} and \textit{singular} otherwise. In the regular case for time-independent Lagrangians the Hamiltonian is the real function $H:T^*Q\rightarrow\mathbb{R}$ defined by $E\circ FL^{-1}$ where $E:TQ\rightarrow \mathbb{R}$ is the familiar energy in the Lagrangian setting. In the singular case the failure of $FL$ to be a diffeomorphism can occur in many ways. A frequent one is that the map is not surjective but defines an embedding of $TQ$ in $T^*Q$. In such situations its image is known as the primary constraint submanifold $\mathsf{M}_0$. When this happens a Hamiltonian can often still be defined as in the regular case, but only on $\mathsf{M}_0$.

In the GNH approach the basic idea is to find the maximal submanifold $\mathsf{M}_1$ of $\mathsf{M}_0$ where the equation $X\iprod \omega=\mathrm{d}H$,  for vector fields $X\in\mathfrak{X}(\mathsf{M}_0)$, has solutions that are tangent to $\mathsf{M}_1$ (here $\omega$ denotes the pull-back of the canonical symplectic form onto the primary constraint submanifold $\mathsf{M}_0$). The previous equation is linear and inhomogeneous so it is possible that it is solvable only if some additional conditions (secondary constraints) hold. Also, if new constraints show up, one has to verify that the vector fields $X$ are also tangent to them.

In practice, when using this method one has to find the constraints, get the equations for the components of the Hamiltonian vector fields and also write the conditions on these components stemming from the consistency of the dynamics (i.e. the requirement that the Hamiltonian vector fields are tangent to the submanifold of $\mathsf{M}_0$ defined by all the secondary constraints). The most direct and natural way to proceed once these equations and consistency conditions are found is to solve the equations and then plug the result into the tangency conditions. Although this can be done, the expressions that one gets are very complicated, to the point that the direct checking of the tangency conditions is almost unfeasible (see \cite{BarberoG:2021ekv,BarberoG:2023qih}). There are, however, three insights that help simplify the computations, solve all the equations and check all the consistency conditions. The first is to look at the form of the gauge transformations of the basic fields (as given by the geometric arguments presented in Section \ref{sec_Cartan_geometry}) and redefine the variables in terms of which the equations of  the Hamiltonian vector fields are written in such a way that the ``trivial'' dynamics corresponding to the gauge transformations is explicitly displayed. When this is done the equations for the Hamiltonian vector fields in terms of the new variables simplify dramatically. A second, helpful insight is to rewrite the consistency conditions also in terms of the new variables. Finally, instead of first solving the equations for the Hamiltonian vector fields and then checking the tangency conditions it is much better to consider all of them at the same time in order to solve them. By proceeding in this way it is actually possible to get the form of the components of the Hamiltonian vector fields in an easily interpretable form. Although we do not give the details in the present work, the result is \textit{exactly} the one presented in Section \ref{sec_Hamiltonians}. Notice that the discussion given in Section \ref{sec_time_gauge} regarding the time gauge and the introduction of the Ashtekar variables remains the same.

Several comments are in order now: the first is that the natural variables to use in the GNH approach applied to the actions that we consider here are the configuration variables in terms of which the Lagrangian is written (which also correspond to the variables used to write the $3+1$ form of the equations in Section \ref{sec_Hamiltonians}). This is so because the primary constraint submanifold $\mathsf{M}_0$ can be naturally parameterized in terms of the variables of the configuration space $Q$. As a consequence the evolution equations (or, equivalently, the Hamiltonian vector fields) that one obtains in the GNH are written also in terms of the configuration variables and, hence, must be equivalent to the ones obtained by the ``projection procedure'' presented in section \ref{sec_Hamiltonians}. Second, notice that the pullback of the canonical symplectic form onto the primary constraint submanifold $\mathsf{M}_0$ can also be written in terms of the configuration variables. As we have seen it is possible to find the way to write this pullback in the usual form in terms of appropriately chosen Darboux coordinates: the Ashtekar variables or the ones corresponding to the SU(2)-ADM formulation from which the Ashtekar formulation can be derived.

Although the Dirac approach is, apparently, very different from the GNH method it shares some features with it. The main steps after identifying the primary constraint submanifold (by proceeding as in the GNH case) is to define the total Hamiltonian $H_T$ by adding the primary constraints to the Hamiltonian with appropriate multipliers and requiring that the dynamics defined by $H_T$ be consistent in the sense that it respects the primary constraints. Generically the equations that fix the value of the multipliers are linear, inhomogeneous equations whose solvability may require the introduction of additional secondary constraints whose dynamical stability must be checked. It is very important to emphasize that the equations for the multipliers must be solved because their solutions, after being introduced in $H_T$, give the final form of the Hamiltonian description of the system dynamics. It is important to bear in mind that the dynamics happens in the submanifold of $\mathsf{M}_0$ defined by the secondary constraints and, hence, the functions representing the constrains as their zero-sets can be used to simplify the final form of the Hamiltonian vector fields.

For the family of actions that we consider in this paper, the equations for the multipliers in $H_T$ together with the conditions enforcing that the Hamiltonian dynamics respects the secondary constraints are exactly the same as those that we have to solve if we rely on the GNH approach and, again, are also exactly the same as those found in Section \ref{sec_Hamiltonians}.

%
%
\section{Conclusions and comments}\label{sec_Conclussions}

In this paper we have discussed a streamlined procedure to get the Hamiltonian description of a number of actions related to GR, in particular the Holst and Cartan-Palatini actions for both Euclidean and Lorentzian GR as well as their Carrollian limit. The procedure presented here takes advantage of several geometric insights gleaned from Cartan geometry which tell us what the relevant geometric objects are and also the form of the gauge symmetries of the actions considered in the paper. Knowing the form of the gauge symmetries is especially useful because it allows us to find the right variables to express the equations for the Hamiltonian vector fields. Although we have not given details on how this would work if the GNH or Dirac methods are used, the geometric insights provided by Cartan geometry are crucial to simplify the, otherwise, long and cumbersome computations. The approach that we have followed is very natural in the context of field theories written in terms of differential forms, in which case the field equations can be interpreted as exterior differential systems (EDS). An interesting study of Dirac's constraints and the GNH procedure in the context of EDS's can be found in \cite{Capriotti:2010}.

There are significant differences between our treatment and the ones in the traditional presentations of the Hamiltonian formulation for gravity in the context of LQG \cite{Ashtekar:1991hf,BarroseSa:2000vx,Alexandrov:2000jw,Peldan:1993hi} and in the Carrollian limit \cite{Sengupta:2022rbd}. First of all, the actions that we consider are written in terms of differential forms (co-tetrads), instead of the associated frames. This has significant advantages in the Hamiltonian treatment because it allows us to use exterior algebra. For instance, the fact that pullbacks interact in a neat way with the exterior differential and the wedge product are crucial to simplify a number of computations. Second, we avoid the use of quadratic constraints because we treat all the variables in the Lagrangian on the same footing (i.e. Lagrange multipliers are considered on the same footing as the rest of the variables). From a conceptual point of view this has some advantages, in particular if the Dirac or GNH approaches are used and prevents conceptual misunderstandings regarding the use of these methods to deal with singular Lagrangian systems. Finally, the fact that we are using first order actions (leading to first order field equations) has two interesting consequences. On one hand the field equations are first order and are written in terms of differential forms. This allows us to write them in a convenient $3+1$ form by using the simple projection procedure described in Section \ref{sec_Hamiltonians}. On the other, since the action is first order the momenta conjugate to the configuration variables are independent of the velocities. As a consequence, the primary constraint submanifold $\mathsf{M}_0$ can be parameterized in terms of the configuration variables only. As a consequence, the evolution will be given just in terms of configuration variables (no momenta) and then it must be equivalent to the one obtained by the projection procedure mentioned above. We conclude then that the best way to find the Hamiltonian description of the dynamics defined by first order actions is the one that we have followed in the paper (which was partially discussed in \cite{Barbero:2021aul}).

An interesting observation is that the self-dual, Euclidean action for GR is also a member of our two-parameter family of actions (it corresponds to $\zeta=+1$ and $\gamma=+1$). However, this is a especial singular case from the perspective of the Hamiltonian analysis \cite{BarberoG:2023qih} that we will study elsewhere with the tools developed in this paper (together with the, harder, Lorentzian case).

%
%
\section*{Acknowledgments}

Some of the ideas that the late Jerzy Lewandowski was exploring in the last years of his life were related to Cartan geometry so we found it fitting to explore if it is possible to employ this beautiful conceptual framework to study the intricacies of the Hamiltonian formulation of gravity. This paper is the result. We thank him wholeheartedly for the inspiration he has given us for so many years.

\begin{appendices}

%
%
%

\section{Notations and all that}\label{app_notation}

The notation used in the paper is chosen to highlight the intrinsic geometric features of the different models used here. Although the readers may feel an initial discomfort or find the notation unfamiliar, we can attest from our own experience that the adaptation to this kind of language pays off handsomely as many computations are not only simple but also very natural. In any case, we provide a dictionary that should allow a smooth transition from the usual language with indices to the more intrinsic one.

All our actions are formulated on an orientable four dimensional manifold $M$. Let $\mathfrak{g}$ be a real, finite-dimensional Lie algebra. The fields that we use to build the actions discussed in the paper are either differential forms of order $p$ (i.e. elements of $\Omega^p(M)$) or $\mathfrak{g}$-valued differential forms (elements of $\Omega^p(M,\mathfrak{g})$). By choosing a linear basis $(X_i)$, $i=1,...,\mathrm{dim}(\mathfrak{g})$ on $\mathfrak{g}$, we can expand a $\mathfrak{g}$-valued $p$-form $\alpha$ as $\alpha=\alpha^i X_i$ with $\alpha^i\in \Omega^p(M)$. In terms of this basis, the Lie bracket can be written with the help of the structure constants $C^i_{\phantom{i}jk}$ as
\[
[X_j,X_k]=C^i_{\phantom{i}jk}X_i\,.
\]
The structure constants satisfy
\begin{align}
  & C^i_{\phantom{i}jk}=-C^i_{\phantom{i}kj}\,,\label{antisymmetry}\\
  & C^\ell_{\phantom{\ell}ij}C^n_{\phantom{n}k\ell}+C^\ell_{\phantom{\ell}jk}C^n_{\phantom{n}i\ell}+C^\ell_{\phantom{\ell}ki}C^n_{\phantom{n}j\ell}=0\,.\label{Jacobi}
\end{align}
All the actions that we consider in the paper are written either in terms of $\mathfrak{su}(2)$-valued differential forms or plain differential forms. As $\mathfrak{su}(2)$ is a semisimple Lie algebra, we can define the non-degenerate and symmetric Cartan-Killing bilinear form $K:\mathfrak{su}(2)\times\mathfrak{su}(2)\rightarrow \mathbb{R}$ which satisfies the invariance property
\begin{equation}\label{CK_invariance}
K(A,K(B,C))=K(K(A,B),C)
\end{equation}
for $A,B,C\in\mathfrak{su}(2)$. In the paper, we rather use the bilinear form $-K/2$, which is also symmetric, non degenerate and invariant in the sense of \eqref{CK_invariance}. We will denote this form as $\langle\cdot,\cdot\rangle$. We can we chose a basis $(X_i)_{i=1,2,3}$ of $\mathfrak{su}(2)$ such that the structure constants are $C^i_{\phantom{i}jk}=\delta^{i\ell}\varepsilon_{\ell jk}$ with $\varepsilon_{\ell jk}$ the Levi-Civita symbol and $\delta^{i\ell}$ the Kronecker delta. In such basis, the bilinear form $\langle\cdot,\cdot\rangle$ is just the Kronecker delta.

\medskip

Let $\alpha\in\Omega^p(\Sigma,\mathfrak{g})$, $\beta\in\Omega^q(\Sigma,\mathfrak{g})$ and expand $\alpha=\alpha^iX_i$ and $\beta=\beta^iX_i$. We introduce now the notation
\[
\begin{array}{l}
\langle \alpha\wed \beta\rangle:=\alpha^i\wedge \beta^j\langle X_i,X_j\rangle\in \Omega^{p+q}(\Sigma)\,,\phantom{\Big|}\\

[\alpha\wed \beta]:=\alpha^i\wedge \beta^j [X_i,X_j]\in \Omega^{p+q}(\Sigma, \mathfrak{g})\,.
\end{array}
\]
If $A\in\Omega^1(\Sigma,\mathfrak{g})$ is a connection 1-form and $\beta\in\Omega^p(\Sigma,\mathfrak{g})$, the covariant exterior differential is defined as
\[
\mathrm{d}_A\beta:=\mathrm{d}\beta+[A\wed \beta]=\mathrm{d}\beta^k X_k+A^i\wedge \beta^j[X_i,X_j]=(\mathrm{d}\beta^k+A^i\wedge \beta^j C_{ij}^{\phantom{ij}k})X_k\,,
\]
and the curvature
\[
F_A:=\mathrm{d}A+\frac{1}{2}[A\wed A]=\mathrm{d}A^kX_k+\frac{1}{2}A^i\wedge A^j[X_i,X_j]=\Big(\mathrm{d}A^k+\frac{1}{2}A^i\wedge A^jC_{ij}^{\phantom{ij}k}\Big)X_k\,.
\]
The following identitites will be used throughout the paper:

\begin{enumerate}
\item Let $\alpha\in\Omega^p(\Sigma,\mathfrak{g})$ and $\beta\in\Omega^q(\Sigma,\mathfrak{g})$, then $\displaystyle\langle \alpha\wed\beta\rangle = (-1)^{pq}\langle\beta\wed\alpha\rangle\,.$
\item Let $\alpha\in\Omega^p(\Sigma,\mathfrak{g})$ and $\beta\in\Omega^q(\Sigma,\mathfrak{g})$, then $\displaystyle[\alpha\wed\beta]= (-1)^{pq+1} [\beta\wed\alpha]\,.$
\end{enumerate}
In many computations, it is necessary to move brackets inside the Cartan-Killing form. The following result is very convenient for this purpose
\begin{enumerate}
\setcounter{enumi}{2}
\item Let $\alpha\in\Omega^p(\Sigma,\mathfrak{g})$, $\beta\in\Omega^q(\Sigma,\mathfrak{g})$ and $\sigma\in\Omega^r(\Sigma,\mathfrak{g})$, then ${\langle[\alpha\wed\beta]\wed\sigma\rangle=\langle\alpha\wed[\beta\wed\sigma]\rangle}$
\end{enumerate}
If, instead, one is interested in moving brackets within brackets the following Jacobi identities for Lie algebra valued forms are helpful
\begin{enumerate}
\setcounter{enumi}{3}
\item Let $\alpha\in\Omega^p(\Sigma,\mathfrak{g})$, $\beta\in\Omega^q(\Sigma,\mathfrak{g})$ and $\sigma\in\Omega^r(\Sigma,\mathfrak{g})$, then
\[
(-1)^{pr}[[\alpha\wed\beta]\wed\sigma]+(-1)^{qp}[[\beta\wed\sigma]\wed\alpha]+(-1)^{rq}[[\sigma\wed\alpha]\wed\beta]=0
\]
\[
(-1)^{pr}[\alpha\wed[\beta\wed\sigma]]+(-1)^{qp}[\beta\wed[\sigma\wed\alpha]]+(-1)^{rq}[\sigma\wed[\alpha\wed\beta]]=0
\]
\end{enumerate}

Finally, we end the appendix with a number of useful identities that we use throughout the paper (here $\iprod$ denotes the interior product of a differential form and a vector field):

\noindent Let $\alpha\in\Omega^p(\Sigma,\mathfrak{g})$, $\beta\in\Omega^q(\Sigma,\mathfrak{g})$ and $X\in\mathfrak{X}(\Sigma)$, then

\begin{enumerate}
\setcounter{enumi}{4}
\item $\displaystyle X\iprod [\alpha\wed\beta]=[(X\iprod\alpha)\wed\beta]+(-1)^p[\alpha\wed (X\iprod\beta)]\,.$
\item $\displaystyle \mathrm{d}_A[\alpha\wed\beta]=[\mathrm{d}_A\alpha\wed\beta]+(-1)^p[\alpha\wed \mathrm{d}_A\beta]\,,$
\item $\displaystyle  \mathrm{d}_A\langle \alpha\wed \beta\rangle=\langle \mathrm{d}_A\alpha\wed \beta\rangle+(-1)^p\langle \alpha\wed \mathrm{d}_A\beta\rangle\,.$
\end{enumerate}
\begin{enumerate}
\setcounter{enumi}{7}
\item Let $\alpha\in\Omega^p(\Sigma,\mathfrak{g})$, then $\mathrm{d}_A^2\alpha=[F_A\wed\alpha]$.
\item $\mathrm{d}_A\hspace*{-1pt}F_A=0$ (Bianchi identity).
\end{enumerate}

%
%
%

\section{Other useful mathematical results}\label{app_math}

\subsection{Simplification of the field equations for the action \eqref{action_Holst}}\label{sub_app_simpli}

Here we show how the equations \eqref{Field_eq_3_bis} and \eqref{Field_eq_2_bis} can be simplified. Checking that \eqref{Field_eq_3_ter} and \eqref{Field_eq_2_ter} imply \eqref{Field_eq_3_bis} and \eqref{Field_eq_2_bis} is a direct exercise. In order to prove the converse we expand
\begin{align}
  &\ddd_A\ee^i=\alpha^i_{\phantom{i}j}\ee^j\wedge\uptau+\beta^i_{\phantom{i}j}\varepsilon^j_{\phantom{j}k\ell}\ee^k\wedge\ee^\ell\,,\label{exp_1}\\
  &\Upomega^i=\Omega^i_{\phantom{i}j}\ee^j+\phi^i\tau\,,\label{exp_2}\\
  &\ddd\uptau=s_i\ee^i\wedge\uptau+m_i\varepsilon^i_{\phantom{i}jk}\ee^j\wedge\ee^k\,.\label{exp_3}
\end{align}
Plugging these into \eqref{Field_eq_3_bis} gives the following relations between the expansion coefficients
\begin{align}
&2m^i+\varepsilon^i_{\phantom{i}jk}\Omega^{jk}=0\,,\label{cond_3}\\
&\varepsilon_{ijk}(\phi^k+s^k)+2\beta_{ij}=0\,.\label{cond_4}
\end{align}
Doing the same with Equation \eqref{Field_eq_2_bis} we obtain the conditions
\begin{align}
  &\varepsilon_{ijk}\beta^{jk}=0\,,\label{cond_1}\\
  &\delta^{ij}\alpha^k_{\phantom{k}k}-\alpha^{ij}+\zeta(\delta^{ij}\Omega^k_{\phantom{k}k}-\Omega^{ij})=0\,.\label{cond_2}
\end{align}
Contracting \eqref{cond_4} with $\varepsilon^{ij\ell}$ and using \eqref{cond_1} one gets $\phi_i+s_i=0$ which, plugged back into \eqref{cond_4} results in $\beta_{ij}=0$.
The trace of \eqref{cond_2} gives $\alpha^k_{\phantom{k}k}+\zeta\Omega^k_{\phantom{k}k}=0$ which, plugged back into  \eqref{cond_2} gives $\alpha_{ij}+\zeta\Omega_{ij}=0$. Gathering all these results for the components of the expansions (\ref{exp_1}-\ref{exp_3}) we immediately arrive at (\ref{Field_eq_3_bis},\ref{Field_eq_2_bis}).

\subsection{Top forms and frame dualities}\label{subsec_topforms}

Let $w$ be a top-form on the 3-manifold $\Sigma$ and consider a reference volume form $\mathsf{vol}_0$. We denote the smooth scalar function $f$ such that $w = f \mathsf{vol}_0$ as
\[
\left(\frac{w}{\mathsf{vol}_0}\right)\,.
\]

Let $e^i \in \Omega^1(\Sigma)$, $i=1,2,3$,  be a non-degenerate spatial cotriad. We define the volume form $\mathsf{vol}_e$ and $\mathsf{det} \, e$ as
\begin{equation}
    \mathsf{vol}_e := \frac{1}{3!}\epsilon_{ijk} e^i \wedge e^j \wedge e^k \,, \quad \mathsf{det}\, e := \left(\frac{\mathsf{vol}_e}{\mathsf{vol}_0}\right) \,.
\end{equation}
Let $E := \frac{1}{2}[e\wed e] \in \Omega^2(\Sigma, \mathfrak{su}(2))$. We introduce the $\gamma$-dependent, ``densitized'' (relative to $\mathsf{vol}_0$) dual vector field $\vec{E}^i \in \mathfrak{X}(\Sigma)$according to:
\begin{equation*}
   \gamma\widetilde{E}^i \iprod \, \mathsf{vol}_0 := E^i \,,
\end{equation*}
that satisfy the condition $\widetilde{E}^i \iprod e_j=\gamma^{-1}(\mathsf{det}e)\delta^i_j$. With the help of $\widetilde{E}$ one finds:
\begin{align}
    e^i &= -\frac{\gamma^2}{2\,\mathsf{det} \, e}\epsilon^{ijk}\widetilde{E}_j \iprod \, \widetilde{E}_k \iprod \, \mathsf{vol}_0 \,, \\
    (\mathsf{det} \, e)^2 &= -\frac{\gamma^3}{3!}\epsilon^{ijk}\widetilde{E}_i \iprod \, \widetilde{E}_j \iprod \, \widetilde{E}_k \iprod \, \mathsf{vol}_0 \,.
\end{align}

\subsection{Computing the Lie derivatives along $\uu$}\label{subsec_Lie}
In order to compute the Lie derivatives along the frame component $\uu$ it is useful to expand
\begin{align*}
&\Big(\FF_\Aa+\frac{\zeta}{2}[\Upomega\wed\Upomega]\Big)^i=\FF^i_{\phantom{i}j}\varepsilon^j_{\phantom{j}j\ell}\ee^k\wedge\ee^\ell+\Phi^{ij}\uptau\wedge\ee_j\,,\\
& \big(\bd_\Aa\Upomega\big)^i=\Ss^i_{\phantom{i}j}\varepsilon^j_{\phantom{j}j\ell}\ee^k\wedge\ee^\ell+\Psi^{ij}\uptau\wedge\ee_j\,.
\end{align*}
The field equations (\ref{Field_eq_1_bis},\ref{Field_eq_4_bis}) and the auxiliary conditions (\ref{c1},\ref{c2}) imply the following conditions on $\FF^{ij}$, $\Phi^{ij}$, $\Ss^{ij}$ and $\Psi^{ij}$:
\begin{align*}
&\FF^i_{\phantom{i}i}=0\,,\quad &&\varepsilon_{ijk}\FF^{jk}=0\,,\quad &&\quad\\
&\Ss^i_{\phantom{i}i}=0\,,\quad &&\varepsilon_{ijk}\Ss^{jk}=0\,,\quad &&\Phi^{ij}=2\zeta\Ss^{ij}\\
&\Phi^i_{\phantom{i}i}=0\,,\quad &&\varepsilon_{ijk}\Phi^{jk}=0\,,\quad &&\Psi^{ij}=2\FF^{ij}\\
&\Psi^i_{\phantom{i}i}=0\,,\quad &&\varepsilon_{ijk}\Psi^{jk}=0\,.\quad &&\quad
\end{align*}
In the computation of $\pounds_\uu\Aa$ and $\pounds_\uu\Aa$ it is necessary to transform two terms that have simple expressions in terms of $\Phi^{ij}$ and $\Psi^{ij}$
\begin{align*}
&\uu\iprod\Big(\FF_\Aa+\frac{\zeta}{2}[\Upomega\wed\Upomega]\Big)=\Phi^{ij}\ee_j\,,\\
&\uu\iprod\big(\bd_\Aa\Upomega\big)^i=\Psi^{ij}\ee_j.
\end{align*}
These, in turn, can be written in a compact form as follows
\begin{align*}
  & \zeta\ast_\ee(\bd_\Aa\Upomega)^i=\zeta\Ss^i_{\phantom{i}j}\varepsilon^{jk\ell}\ast_\ee(\ee_k\wedge\ee_\ell)=2\zeta \Ss^{ij}\ee_j=\Phi^{ij}e_j=\uu\iprod\Big(\FF_\Aa+\frac{\zeta}{2}[\Upomega\wed\Upomega]\Big)\,,\\
 & \zeta\ast_\ee\Big(\FF_\Aa+\frac{\zeta}{2}[\Upomega\wed\Upomega]\Big)^i=\FF^i_{\phantom{i}j}\varepsilon^{jk\ell}\ast_\ee(\ee_k\wedge\ee_\ell)
 =2\FF^{ij}\ee_j=\Psi^{ij}e_j=\uu\iprod\big(\bd_\Aa\Upomega\big)^i\,,
 \end{align*}
where $\ast_\ee$ is defined in such a way that $\ast_\ee(\ee_k\wedge\ee_\ell)=\varepsilon_{k\ell m}\ee^m$.

\subsection{Hodge duals}\label{subsec_Hodge}

If we have a co-frame $e$ on a 3-manifold $\Sigma$ the Hodge duals, denoted by $\ast$ are
\begin{align*}
  & * 1=\mathsf{vol}_e\,, \\
  & * e=\frac{1}{2}[e\wed e]\,, \\
  & * \frac{1}{2}[e\wed e]=e\,, \\
  & * \mathsf{vol}_e=1\,.
\end{align*}
If we label the internal components of the coframe as $e_i$ ($i=1,2,3$) the preceding definitions are equivalent to
\begin{align*}
  & \quad\,\,\,\,\,\ast 1=\mathsf{vol}_e\,, \\
  & \quad\,\,\,\,\,\ast e_i=\frac{1}{2}\varepsilon_i^{\phantom{i}jk}e_j\wedge e_k\,, \\
  & \quad\,\,\,\,\,\ast (e_i\wedge e_j)=\varepsilon_{ij}^{\phantom{ij}k}e_k\,, \\
  & \quad\,\,\,\,\,\ast \mathsf{vol}_e=1\,.
\end{align*}

\subsection{Duals and $\mathfrak{su}(2)$ valued 2-forms}\label{subsec_duals}

In order to write the evolution vector fields in an intrinsic geometric way the Hodge duals come in handy. Let $B\in\Omega^2(\Sigma,\mathfrak{su}(2))$ and $J_i$ the basis of $\mathfrak{su}(2)$ generators introduced above and $\tau\in\Omega^1(\Sigma)\,,\tau=\tau^i e_i$, we can then expand $B$ as
\[
B=\frac{1}{2}\mathsf{B}_{ij}^{\phantom{ij}k}e^i\wedge e^j J_k=\frac{1}{2}\mathsf{B}^{k\ell}\varepsilon_{\ell ij}e^i\wedge e^j J_k\,.
\]
The following identities are then easy to prove
\begin{align}
&(\ast \!B)^i=\mathsf{B}^{ij}e_j\,,\\
&(\ast(\tau\wedge\ast B))^i=-\mathsf{B}^{ij}\tau^k\varepsilon_{jk\ell}e^\ell\,,\\
&\mathsf{B}^{ij}\varepsilon_{ijk}=\mathsf{B}_{ki}^{\phantom{ki}i}=-\ast(\langle\ast B\wed e\rangle\wedge e_k)\,.
\end{align}

\end{appendices}

%
%
\section*{Declarations}

\noindent \textbf{Competing interests:} The authors declare no competing interests.

\medskip

\noindent \textbf{Funding:} F. Barbero and E. J. S. Villase\~nor are supported by the Spanish Ministerio de Ciencia Innovaci\'on y Universidades-Agencia Estatal de Investigaci\'on grant PID2025-168050NB-I00. E. J. S. Villase\~nor and V. Varo are supported by the Spanish Ministerio de Ciencia Innovaci\'on y Universidades-Agencia Estatal de Investigaci\'on grant PID2025-172146NB-I00. J. Margalef-Bentabol is supported by NSERC grant DGECR-2026-00316. F. Barbero acknowledges the support of the WOST, WithOut SpaceTime project (https://withoutspacetime.org), supported by Grant ID\# 63683 from the John Templeton Foundation (JTF). The opinions expressed in this work are those of the authors and do not necessarily reflect the views of the John Templeton Foundation.

%
%


\begin{thebibliography}{99}

\bibitem{Dirac1} P. A. M. Dirac, \textit{The Theory of Gravitation in Hamiltonian Form}, \href{https://doi.org/10.1098/rspa.1958.0142}{Proc. R. Soc. Lond. A Math. Phys. Sci. 246 (1958) 333}

\bibitem{Arnowitt:1962} R. L. Arnowitt, S. Deser and C. W. Misner, \textit{The Dynamics of general relativity}, in \textit{Gravitation: An Introduction to Current Research}, L. Witten Ed., John Wiley \& Sons (New York, London), (1962) Chapter 7, 227–265.

\bibitem{Arnowitt:1962hi} R. L. Arnowitt, S. Deser and C. W. Misner, \textit{The Dynamics of general relativity}, \href{https://doi.org/10.1007/s10714-008-0661-1}{Gen. Rel. Grav. 40 (2008) 1997}

\bibitem{Ashtekar:1986yd} A. Ashtekar, \textit{New Variables for Classical and Quantum Gravity}, \href{https://doi.org/10.1103/PhysRevLett.57.2244}{Phys. Rev. Lett. 57 (1986) 2244}

\bibitem{Ashtekar:1987gu} A. Ashtekar, \textit{New Hamiltonian Formulation of General Relativity}, \href{https://doi.org/10.1103/PhysRevD.36.1587}{Phys. Rev. D 36 (1987) 1587}

\bibitem{Dirac2} P. A. M. Dirac \textit{Lectures on Quantum Mechanics} (New York: Dover) 2001.

\bibitem{BarberoG:2019ajq} J. F. Barbero G., B. D\'{\i}az, J. Margalef-Bentabol and E. J. S. Villase\~nor, \textit{Dirac{\textquoteright}s algorithm in the presence of boundaries: a practical guide to a geometric approach}, \href{https://doi.org/10.1088/1361-6382/ab436b}{Class. Quant. Grav. 36 (2019) 205014}

\bibitem{Gotay1} M. J. Gotay, J. M. Nester and G. Hinds, \textit{Presymplectic Manifolds and the Dirac–Bergmann Theory of Constraints}, \href{https://doi.org/10.1063/1.523597}{J. Math. Phys. 19 (1978) 2388}

\bibitem{Gotay2} M. J. Gotay, \textit{Presymplectic Manifolds, Geometric Constraint Theory and the Dirac-Bergmann Theory of Constraints,} Ph.D. thesis, Center for Theoretical Physics of the University of Maryland, 1979.

\bibitem{Wise:2009fu} D. K, Wise, \textit{Symmetric space Cartan connections and gravity in three and four dimensions}, \href{https://doi.org/10.3842/SIGMA.2009.080}{SIGMA 5 (2009) 080}

\bibitem{Wise:2006sm} D. K. Wise, \textit{MacDowell-Mansouri gravity and Cartan geometry}, \href{https://doi.org/10.1088/0264-9381/27/15/155010}{Class. Quant. Grav. 27 (2010) 155010}

\bibitem{sharpe2000differential} R. ~W. Sharpe, \emph{Differential geometry: Cartan's generalization of Klein's Erlangen program}, vol. 166, Springer Science \& Business Media, 2000.

\bibitem{eschenburg1997lecture} J.-H. Eschenburg \textit{Lecture Notes on Symmetric Spaces}, \href{https://myweb.rz.uni-augsburg.de/~eschenbu/symspace.pdf}{University of Augsburg}

\bibitem{Freed:1992vw} D. S. Freed, \textit{Classical Chern-Simons theory. Part 1}, \href{https://doi.org/10.1006/aima.1995.1039}{Adv. Math. 113 (1995) 237}

\bibitem{Tejero:2004} C. Tejero Prieto, \textit{Variational formulation of Chern-Simons theory for arbitrary Lie groups}, \href{https://doi.org/10.1016/j.geomphys.2003.11.005}{J. Geom. Phys. 50 (2004) 138}

\bibitem{Figueroa-OFarrill:2022mcy} J. Figueroa-O'Farrill, E. Have, S. Prohazka, and J. Salzer, \textit{The gauging procedure and carrollian gravity}, \href{https://link.springer.com/article/10.1007/JHEP09(2022)243}{JHEP, 09 (2022) 243}

\bibitem{Margalef-Bentabol:2020teu} J. Margalef-Bentabol and E. J. S. Villase\~nor, \textit{Geometric formulation of the Covariant Phase Space methods with boundaries}, \href{https://doi.org/10.1103/PhysRevD.103.025011}{Phys. Rev. D 103 (2021) 025011}

\bibitem{Margalef-Bentabol:2022zso} J. Margalef-Bentabol and E. J. S. Villase\~nor, \textit{Proof of the equivalence of the symplectic forms derived from the canonical and the covariant phase space formalisms}, \href{https://doi.org/10.1103/PhysRevD.105.L101701}{Phys. Rev. D 105 (2022) L101701}

\bibitem{Margalef-Bentabol:2018} J. Margalef-Bentabol, \textit{Towards general relativity through parametrized theories}, \href{https://doi.org/10.48550/arXiv.1807.05534}{Ph.D. thesis (2018), Universidad Carlos III de Madrid}

\bibitem{BarroseSa:2000vx} N. Barros e S\'a, \textit{Hamiltonian analysis of general relativity with the Immirzi parameter}, \href{https://doi.org/10.1142/S0218271801000858}{Int. J. Mod. Phys. D 10 (2001) 261}

\bibitem{Alexandrov:2000jw} S. Alexandrov, \textit{$SO(4,\mathbb{C})$ covariant Ashtekar-Barbero gravity and the Immirzi parameter}, \href{https://doi.org/10.1088/0264-9381/17/20/307}{Class. Quant. Grav. 17 (2000) 4255}

\bibitem{BarberoG:2020tit} J. F. Barbero G., B. D\'{\i}az, J. Margalef-Bentabol and E. J. S. Villase\~nor, \textit{Concise symplectic formulation for tetrad gravity}, \href{https://doi.org/10.1103/PhysRevD.103.024051}{Phys. Rev. D 103 (2021) 024051}

\bibitem{BarberoG:2026urv} J. F. Barbero G, J. Margalef-Bentabol, V. Varo and E. J. S. Villase{\~n}or, \textit{From Holst to Carroll Gravity: a Hamiltonian point of view}, \href{https://doi.org/10.1103/cph1-yj1s}{Phys. Rev. D 114 (2026) 065024}

\bibitem{Barbero:2021aul} J. F. Barbero G, M. Basquens, V. Varo and E. J. S. Villase{\~n}or, \textit{Three Roads to the Geometric Constraint Formulation of Gravitational Theories with Boundaries}, \href{https://doi.org/10.3390/sym13081430}{Symmetry 13 (2021) 1430}

\bibitem{BarberoG:2021ekv} J. F. Barbero G., B. D\'{\i}az, J. Margalef-Bentabol and E. J. S. Villase\~nor, \textit{Hamiltonian Gotay-Nester-Hinds analysis of the parametrized unimodular extension of the Holst action}, \href{https://doi.org/10.1103/PhysRevD.103.064062}{Phys. Rev. D 103 (2021) 064062}

\bibitem{BarberoG:2023qih} J. F. Barbero G., M. Basquens and E. J. S. Villase{\~n}or, \textit{Euclidean self-dual gravity: Ashtekar variables without gauge fixing}, \href{https://doi.org/10.1103/PhysRevD.109.064047}{Phys. Rev. D 109 (2024) 064047}

\bibitem{Capriotti:2010} S. Capriotti, \textit{Dirac constraints in field theoryand  exterior differential systems}, \href{https://doi.org/10.3934/jgm.2010.2.1}{J. Geom. Mech. 2(1) (2010) 1}

\bibitem{Ashtekar:1991hf} A. Ashtekar, \textit{Lectures on nonperturbative canonical gravity}, \href{https://doi.org/10.1142/1321}{World Scientific (1991)}


\bibitem{Peldan:1993hi} P. Peld\'an, \textit{Actions for gravity, with generalizations: A Review}, \href{https://doi.org/10.1088/0264-9381/11/5/003}{Class. Quant. Grav. 11 (1994) 1087}

\bibitem{Sengupta:2022rbd}  S. Sengupta, \textit{Hamiltonian form of Carroll gravity},   \href{https://doi.org/10.1103/PhysRevD.107.024010}{Phys. Rev. D 107 (2023) 024010}


\end{thebibliography}
\end{document}